\pdfoutput=1

\documentclass[11pt,twoside,a4paper,cmspaper,final,collab]{cms-tdr}
\def\svnVersion{374485}\def\cmsCernNoTag{CERN-PH-EP/2015-063}\def\cmsCernDate{\today}\def\cmsMessage{Published in the Journal of High Energy Physics as \href{http://dx.doi.org/10.1007/JHEP07(2015)042}{\doi{10.1007/JHEP07(2015)042.}}}

\begin{document}\cmsNoteHeader{EXO-14-008}

\hyphenation{had-ron-i-za-tion}
\hyphenation{cal-or-i-me-ter}
\hyphenation{de-vices}
\RCS$Revision: 291426 $
\RCS$HeadURL: svn+ssh://svn.cern.ch/reps/tdr2/papers/EXO-14-008/trunk/EXO-14-008.tex $
\RCS$Id: EXO-14-008.tex 291426 2015-06-05 13:37:17Z peiffer $
\providecommand{\tauh}{\ensuremath{\tau_\mathrm{h}}\xspace}
\providecommand{\ST}{\ensuremath{S_\mathrm{T}}\xspace}
\providecommand{\LQt}{\ensuremath{\mathrm{LQ}_3}\xspace}
\providecommand{\ALQt}{\ensuremath{\overline{\mathrm{LQ}}_3}\xspace}
\cmsNoteHeader{EXO-14-008}
\title{Search for third-generation scalar leptoquarks in the t$\tau$ channel in proton-proton collisions at $\sqrt{s}=8$\TeV}

\date{\today}

\abstract{
A search for pair production of third-generation scalar leptoquarks decaying to top quark and $\tau$ lepton pairs is presented using proton-proton collision data at a center-of-mass energy of $\sqrt{s}=8$\TeV collected with the CMS detector at the LHC and corresponding to an integrated luminosity of 19.7\fbinv.
The search is performed using events that contain an electron or a muon, a hadronically decaying $\tau$ lepton, and two or more jets.
The observations are found to be consistent with the standard model predictions.
Assuming that all leptoquarks decay to a top quark and a $\tau$ lepton, the existence of pair produced, charge $-1/3$, third-generation leptoquarks up to a mass of 685\GeV is excluded at 95\% confidence level.
This result constitutes the first direct limit for leptoquarks decaying into a top quark and a $\tau$ lepton, and may also be applied directly to the pair production of bottom squarks decaying predominantly via the R-parity violating coupling $\lambda^{\prime}_{333}$.
}

\hypersetup{%
pdfauthor={CMS Collaboration},%
pdftitle={Search for third-generation scalar leptoquarks in the t-tau channel in proton-proton collisions at 8 TeV},
pdfsubject={CMS},%
pdfkeywords={CMS, physics, leptoquarks, same-sign dileptons, third generation}}

\maketitle

\section{Introduction}

Leptoquarks (LQ) are hypothetical particles that carry both lepton (L) and baryon (B) quantum numbers. They appear in theories beyond the standard model (SM), such as grand unification~\cite{PatiSalam,GeorgiGlashow,TOPSU(5)Model}, technicolor~\cite{Technicolor}, and compositeness~\cite{LightLeptoquarks,Gripaios:2009dq} models.
A minimal extension of the SM to include all renormalizable gauge invariant interactions, while respecting existing bounds from low-energy and precision measurements leads to the effective Buchm\"{u}ller--R\"{u}ckl--Wyler model~\cite{Buchmuller1987442}.
In this model, LQs are assumed to couple to one generation of chiral fermions, and to separately conserve L and B quantum numbers.
An LQ can be either a scalar (spin 0) or a vector (spin 1) particle with a fractional electric charge.
A comprehensive list of other possible quantum number assignments for LQs coupling to SM fermions can be found in Ref.~\cite{LQsearchesAtHeraTevatron}.

This paper presents the first search for a third-generation scalar LQ (\LQt) decaying into a top quark and a $\tau$ lepton.
Previous searches at hadron colliders have targeted LQs decaying into quarks and leptons of the first and second generations~\cite{Aad:2011ch,ATLAS:2012aq,Chatrchyan:2012vza} or
the third-generation in the $\LQt \to \PQb \nu$ and $\LQt \to \PQb \tau$ decay channels~\cite{PhysRevLett.99.061801,PhysRevD.77.091105,ATLAS:2013oea,Chatrchyan:2012st,Khachatryan:2014ura,Khachatryan:2015wza}.
The presented search for third-generation LQs can also be interpreted in the context of R-parity violating (RPV) supersymmetric models~\cite{Barbier:2004ez}
where the supersymmetric partner of the bottom quark (bottom squark) decays into a top quark and a $\tau$ lepton via the RPV coupling $\lambda^{\prime}_{333}$.

At hadron colliders, such as the CERN LHC, LQs are mainly pair produced through the quantum chromodynamic (QCD) quark-antiquark annihilation and gluon-gluon fusion sub\-pro\-ces\-ses.
There is also a lepton mediated $t$($u$)-channel contribution that depends on the unknown lepton-quark-LQ Yukawa coupling, but this contribution is suppressed at the LHC for the production of third-generation LQs as it requires third-generation quarks in the initial state.
Hence, the LQ pair production cross section depends only upon the assumed values of the LQ spin and mass, and upon the proton-proton center-of-mass energy.
We consider scalar LQs in the mass range up to several hundred \GeVns{}.
The corresponding next-to-leading-order (NLO) pair production cross sections and associated uncertainties at $\sqrt{s}=8$\TeV are taken from the calculation presented in Ref.~\cite{Kramer}.

It is customary to denote the branching fractions of LQs into a quark and a charged lepton or a quark and a neutrino within the same generation as $\beta$ and $1-\beta$, respectively.
Assuming that third-generation scalar LQs with charge $-1/3$ exclusively couple to quarks and leptons of the third-generation, the two possible decay channels are $\LQt \to \PQt \tau^{-}$ and $\LQt \to \PQb \nu$.
In this paper, we initially assume that $\beta=1$ so that the $\LQt$ always decays to a t$\tau$ pair. The results are then reinterpreted as a function of the branching
fraction with $\beta$ treated as a free parameter.

\begin{figure}[tp]
\centering
\includegraphics[width=\textwidth]{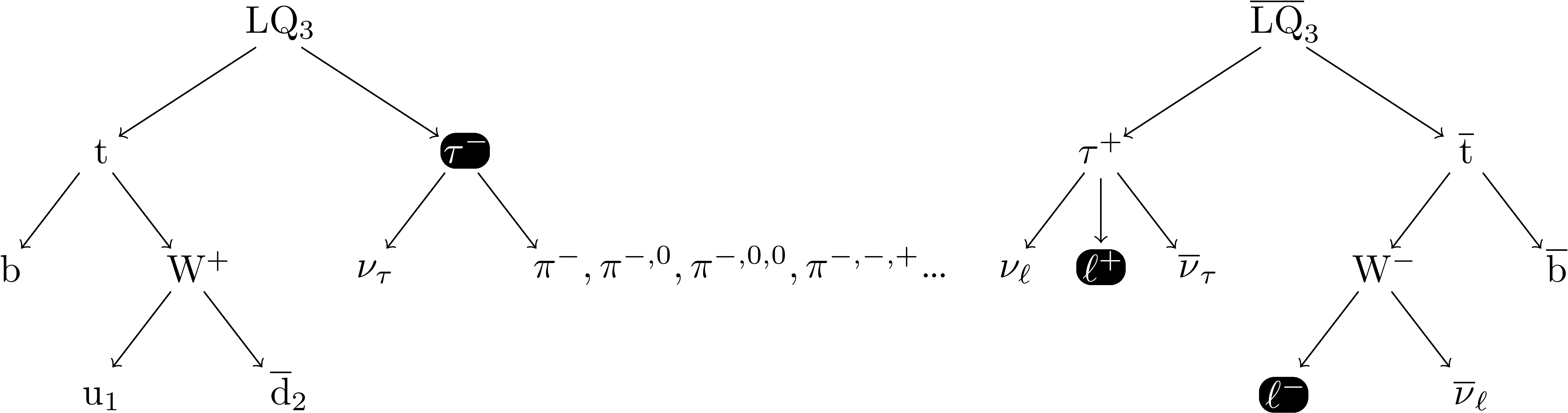}
\caption{One of the $\LQt\ALQt$ decay chains with both same-sign and opposite-sign $\ell\tauh$ pairs. Labels u and d denote up and down type quarks, and $\ell$ denotes an electron or a muon.\label{lqDecayChainPlot}}
\end{figure}

We consider events with at least one electron or muon and one $\tau$ lepton where the $\tau$ lepton undergoes a one- or three-prong hadronic decay, $\tauh \to \text{hadron(s)}+\nu_\tau$.
In $\LQt\ALQt$ decays, $\tau$ leptons arise directly from LQ decays, as well as from W bosons in the top quark decay chain, whereas electrons and muons are produced only in leptonic decays of W bosons or $\tau$ leptons.
The major backgrounds come from $\ttbar$+jets, Drell--Yan(DY)+jets, and W+jets production,
where a significant number of events have jets misidentified as hadronically decaying $\tau$ leptons.
The search is conducted in two orthogonal selections, labelled as category A and category B.
In category A, a same-sign $\mu\tauh$ pair is required in each event, which suppresses SM backgrounds.
Misidentified $\tauh$ candidates originating from jets constitute the main background in category A.
Category B utilizes both $\Pe\tauh$ and $\mu\tauh$ pairs with slightly relaxed $\tau$ lepton identification criteria without imposing a charge requirement on the lepton pair.
This yields a higher signal acceptance, but a larger irreducible background from SM processes.
In order to keep the two samples statistically independent, events that satisfy the category A criteria are removed from the category B sample.
Figure~\ref{lqDecayChainPlot} shows a schematic representation of an $\LQt\ALQt$ decay chain that can satisfy the requirements for both categories.

The signature for this search is chosen to be $\ell\tauh$+$X$, where $\ell$ denotes an electron or a muon, and $X$ is two or more jets and any additional charged leptons in category A, or three or more jets and any additional charged leptons in category B.
The additional jet requirement in category B is beneficial in suppressing background events from dominant SM processes with two jets and an opposite-sign $\ell\tauh$ pair.

\section{Reconstruction and identification of physics objects}

The CMS apparatus is a multipurpose particle detector with a superconducting solenoid of 6\unit{m} internal diameter, which provides a magnetic field of 3.8\unit{T}.
Within the volume of the solenoid are a silicon pixel and strip tracker, a lead tungstate crystal electromagnetic calorimeter, and a brass and scintillator hadron calorimeter, each composed of a barrel and two endcap sections.
Muons are measured in gas-ionization detectors embedded in the steel flux-return yoke outside the solenoid.
Extensive forward calorimetry complements the coverage provided by the barrel and endcap detectors.
A more detailed description of the CMS detector, together with a definition of the coordinate system used and the relevant kinematic variables, can be found in Ref.~\cite{CMSdetector}.

The electron, muon, and $\tau$ lepton candidates used in this paper are reconstructed using a particle-flow (PF) event reconstruction technique~\cite{CMS:2009nxa,CMS-PAS-PFT-10-001}
which reconstructs and identifies single particles (muons, electrons, charged/neutral hadrons, and photons) using an optimized combination of all subdetector information.

Muon candidates are reconstructed from a combined track in the muon system and the tracking system \cite{Chatrchyan:2012xi}.
The hadronically decaying $\tau$ lepton candidates are reconstructed via the ``hadron-plus-strips'' algorithm which combines one or three charged hadrons with up to two neutral pions
that are reconstructed from PF candidates combining tracker and calorimeter information~\cite{CMS-TAU-paper}.
Electron candidates are obtained by reconstructing trajectories from hits in the tracker layers and energy depositions in the electromagnetic calorimeter with a Gaussian sum filter~\cite{Khachatryan:2015hwa}.

Jets are reconstructed by using the anti-\kt algorithm~\cite{CMS-PAS-PFT-10-001,antiKT,JEScalib} to cluster PF candidates with a distance parameter of $\Delta R=0.5$ (where $\Delta R=\sqrt{\smash[b]{(\Delta\eta)^2+(\Delta\phi)^2}}$, $\eta$ denotes the pseudorapidity and $\phi$ denotes the azimuthal angle in radians).
The missing transverse momentum $\ptvecmiss$ is calculated as a negative vectorial sum of the transverse momenta of all the PF candidates.
The missing transverse energy \MET is defined as the magnitude of the $\ptvecmiss$ vector.
Jet energy corrections are applied to all jets and are also propagated to the calculation of \ETmiss~\cite{CMS-PAS-JME-12-002}.

The collisions are selected using a two-tiered trigger system, composed of a hardware based level-1 trigger and a software based high-level trigger (HLT)~\cite{Adam:2005zf} running on a computing farm.

The following quantities are constructed using the physics objects described earlier:

\begin{itemize}
\item $\ST$ is the scalar $\pt$ sum of all objects in the event, including muons, hadronically decaying $\tau$ leptons, electrons, jets, and \ETmiss.
\item $M_\mathrm{T}(\ell,\ptvecmiss)$ is the transverse mass, $\sqrt{2\pt^{\ell}\ETmiss(1-\cos(\Delta\phi(\ptvecmiss,\ell)))}$, reconstructed from the given lepton and the $\ptvecmiss$ in the event where $\Delta\phi(\ptvecmiss,\ell)$ is the difference in the azimuthal angle between the directions of the missing transverse momentum and the lepton momentum.
\item $\widetilde{\abs{\eta}}$ is the pseudorapidity defined as $\widetilde{\abs{\eta}} = -\ln{\tan{(\bar{\theta}/2)}}$, where $\bar{\theta}$ is the average absolute polar angle of all electrons, muons, and hadronically decaying $\tau$ leptons in an event as measured from the beam-axis in the lab frame, and is used as a measure of the event centrality.
\end{itemize}

\section{Data and simulated samples \label{DataMCsamples}}

This analysis uses data collected with the CMS detector at the LHC during proton-proton (pp) collisions at $\sqrt{s}=8$\TeV.
Proton bunches were separated by 50\unit{ns} and the average number of additional primary vertices in the collision of the two beams in the same proton bunch crossing was 20 (pileup).
The search is conducted using a combination of isolated and non-isolated single-muon data corresponding to an integrated luminosity of 19.5\fbinv in category A, and using isolated single-muon or single-electron data corresponding to an integrated luminosity of 19.7\fbinv in category B.
The muon triggers require a muon candidate to have $\pt>24$\GeV and $\abs{\eta}<2.1$.
The electron trigger requires an isolated electron candidate with $\pt>27$\GeV and $\abs{\eta}<2.5$.

The LQ signal processes have been simulated using the \PYTHIA generator (v6.426)~\cite{pythia}.
Single top quark and top quark pair production have been simulated with \POWHEG (v1.0)~\cite{powheg1,Alioli:2009je,Re:2010bp,Frixione:2007nw}.
For the W+jets background, DY+jets processes, and \ttbar production in association with W or Z bosons, \MADGRAPH (v5.1) has been used~\cite{madgraph}.
Diboson and QCD multijet processes as well as processes involving Higgs bosons have been generated with \PYTHIA, other SM backgrounds have been simulated with \MADGRAPH.
The parton shower and hadronization in samples generated with \POWHEG or \MADGRAPH has been performed with \PYTHIA. In case of \MADGRAPH, the matching to \PYTHIA has been done with the MLM scheme~\cite{MLM}.
In all of the generated samples, $\tau$ lepton decays were simulated via \TAUOLA \cite{tauola} and the response of the CMS detector has been simulated with \GEANTfour~\cite{geant4}.
The \POWHEG samples are produced with the CT10~\cite{Lai:2010vv} parton distribution function (PDF), all other samples have been generated using CTEQ6L1~\cite{CTEQpaper} PDF set.
The Monte Carlo (MC) simulated events are re-weighted to account for differences in trigger and lepton reconstruction efficiencies, pileup modeling, and  jet/missing transverse energy response of the detector.
The simulated events are normalized using next-to-next-to-leading-order (NNLO) (W+jets, DY+jets~\cite{Li:2012wna}, $\ttbar$+jets~\cite{Czakon:2013goa}, WH, ZH~\cite{Brein:2012ne}), approximate NNLO (t, tW~\cite{Kidonakis:2008mu}), NLO (diboson~\cite{Campbell:2011bn}, $\ttbar\Z$~\cite{ttWZ}, $\ttbar\PW$~\cite{ttWZ,ttW}, $\ttbar\PH$~\cite{Frederix:2011zi,Garzelli:2011vp}, triboson~\cite{Alwall:2014hca}), or leading-order ($\PW^\pm\PW^\pm\PQq\PQq$, $\ttbar\PW\PW$, $\PW\gamma^{*}$, QCD multijet~\cite{pythia,madgraph}) cross sections at $\sqrt{s}=8$\TeV.

The characteristics of the simulated $\ttbar$+jets and W+jets events have been found to contain discrepancies when compared with measurements of the $\pt$ spectrum of top quarks~\cite{CMS-PAS-TOP-12-028} and the leading jet~\cite{Khachatryan:2014uva}, respectively.
Re-weighting factors, parametrized as functions of the respective $\pt$ distributions, are applied to the simulated events to correct for these discrepancies.
The correction factors for $\ttbar$+jets~\cite{CMS-PAS-TOP-12-028} range up to 30\% whereas the correction factors for the W+jets samples vary between 8\% and 12\%.

\section{Event selection\label{Selection}}

A summary of the search regions, selection criteria, and the methods used to determine background contributions for categories A and B is given in Table~\ref{SelScheme}.

\begin{table}
\centering
{\caption{Summary of the search strategies in event categories A and B. }\label{SelScheme}}
\resizebox{\linewidth}{!}{%
\begin{tabular}{ l |  l  l  }
\hline
&  Category A & Category B \\ \hline
\multirow{1}{*}{Lepton selection} & Same-sign $\mu\tauh$ pair &  $\mu\tauh$ or $\Pe\tauh$ pair \\
& & (category A events are removed) \\[1.2ex]
Jet selection & At least two jets & At least three jets\\[1.2ex]
$\ETmiss$ requirement & No $\ETmiss$ requirement & $\ETmiss>50$\GeV \\[1.2ex]
$\ST$ and $\tau$ lepton $\pt$ & Optimized for each LQ  & $\ST>1000$\GeV, $\pt^{\tau}>20$\GeV \\
 requirements  & mass hypothesis & \\[1.2ex]
\multirow{3}{*}{Background estimation} & Main component containing        & Estimated via simulation, corrections\\
                                       & misidentified muons $\&$ $\tau$ leptons & applied for $\tau$ lepton misidentification \\
                                       & estimated using data events    & rate and top quark and W $\pt$ distributions\\[1.2ex]
\multirow{2}{*}{Search regions} & \multirow{2}{*}{2 search regions binned in $\widetilde{\abs{\eta}}$} & 8 search regions in 4 $\tau$ lepton $\pt$ regions \\
                                 &                                                                  & for $\mu\tauh$ and $\Pe\tauh$ channels \\[1.2ex]
\hline
\end{tabular}
}
\end{table}

\subsection{Event selection in category A\label{SelectionCatA}}

In category A,
two selections, denoted as loose and tight, are defined for the muon and $\tau$ lepton candidates, which differ only in the thresholds of the isolation requirements.
The tight selections are applied to define the signal region, and the loose selections are used in the estimation of backgrounds as defined in Section~\ref{BackgroundsCatA}.

Muon candidates are required to have $\pt>25$\GeV and $\abs{\eta}<2.1$.
The loose muon selection has no isolation requirement, whereas the tight muon selection requires the scalar $\pt$ sum of all PF candidates in a cone of radius $\Delta R=0.4$ around the muon to be less than 12\% of the muon $\pt$.
The muon kinematic and isolation thresholds are chosen to match the trigger requirements used in selecting the events.

Hadronically decaying $\tau$ lepton candidates are required to satisfy $\pt>20$\GeV and $\abs{\eta}<2.1$.
For the loose $\tau$ lepton selection, the scalar $\pt$ sum of charged hadron and photon PF candidates with $\pt>0.5$\GeV in a cone of radius $\Delta R=0.3$ around the $\tau$ lepton candidate is required to be less than 3\GeV.
For the tight $\tau$ lepton selection, a more restrictive isolation requirement is applied with a cone of radius $\Delta R=0.5$ and a threshold value of 0.8\GeV~\cite{CMS-TAU-paper}.
All $\tauh$ candidates are required to satisfy a requirement that suppresses the misidentification of electrons and muons as hadronically decaying $\tau$ leptons~\cite{CMS-TAU-paper}.

Electron candidates are required to have $\pt>15$\GeV and $\abs{\eta}<2.5$.
The ratio of the scalar $\pt$ sum of all PF candidates in a cone of radius $\Delta R=0.3$ around the electron object, relative to the electron $\pt$, is required to be less than 15\%.

All muon, electron, and $\tauh$ candidates are required to be separated by $\Delta R>0.3$ from each other.
In addition, the separation between the muon and $\tau$ lepton candidates and the nearest jet to which they do not contribute is required to be $\Delta R(\mu,j)_{\text{min}}>0.5$ and $\Delta R(\tau,j)_{\text{min}}>0.7$ respectively. This requirement is imposed in order to reduce the impact of QCD jet activity on the respective isolation cones.

Jet candidates are required to have $\pt>40$\GeV, $\abs{\eta}<3$. Jets overlapping with the electron, muon, and $\tauh$ candidates
within a cone of $\Delta R=0.5$ are not considered.

Each event is required to contain a same-sign $\mu\tauh$ pair, chosen among the muon and $\tauh$ lepton candidates satisfying the loose selection criteria.
If the event contains more than one $\mu\tauh$ pair, the same-sign pair with the largest scalar sum $\pt$ is selected.
The selected $\mu\tauh$ pair is then required to satisfy the tight selection criteria.
Events failing the tight selection criteria on one or both leptons are utilized in the estimation of backgrounds described in Section \ref{BackgroundsCatA}.

For the signal selection, same-sign $\mu\tauh$ events  are required to have $\ST>$ 400\GeV and two or more jets.
Events containing an opposite-sign dimuon pair with an invariant mass within 10\% of the Z boson mass are vetoed.
In order to exploit a feature of the signal model that produces the $\LQt$ pair dominantly in the central region, the search is split into two channels with $\widetilde{\abs{\eta}}<0.9$ (central) and $\widetilde{\abs{\eta}}\ge0.9$ (forward).
Furthermore, a 2D optimization is performed using the simulated samples for the determination of selection criteria in the ($\ST$,$\pt^{\tau}$) plane for each $\LQt$ mass hypothesis in the range of 200--800\GeV.
The $\pt^{\tau}$ requirement is only applied to the $\tau$ lepton candidate that is a part of the selected same-sign $\mu\tauh$ pair.
The optimization is accomplished by maximizing the figure of merit given in Eq.~(\ref{PunziEqu})~\cite{Punzi:2003bu}:
\begin{equation}\label{PunziEqu}
\chi(\pt^{\tau},\ST)=\frac{\varepsilon(\pt^{\tau},\ST)}{1+\sqrt{B(\pt^{\tau},\ST)}}
\end{equation}
where $\varepsilon$ is the signal efficiency and $B$ is the number of background events.
The ($\ST$,$\pt^{\tau}$) thresholds have been optimized in the central channel and applied identically in the forward channel. These optimized selections and the corresponding efficiencies as a function of the $\rm{LQ}_3$ mass are presented later, in Section~\ref{ResultsSection}.

\begin{figure}[htp]
\centering
\includegraphics[width=0.49\textwidth]{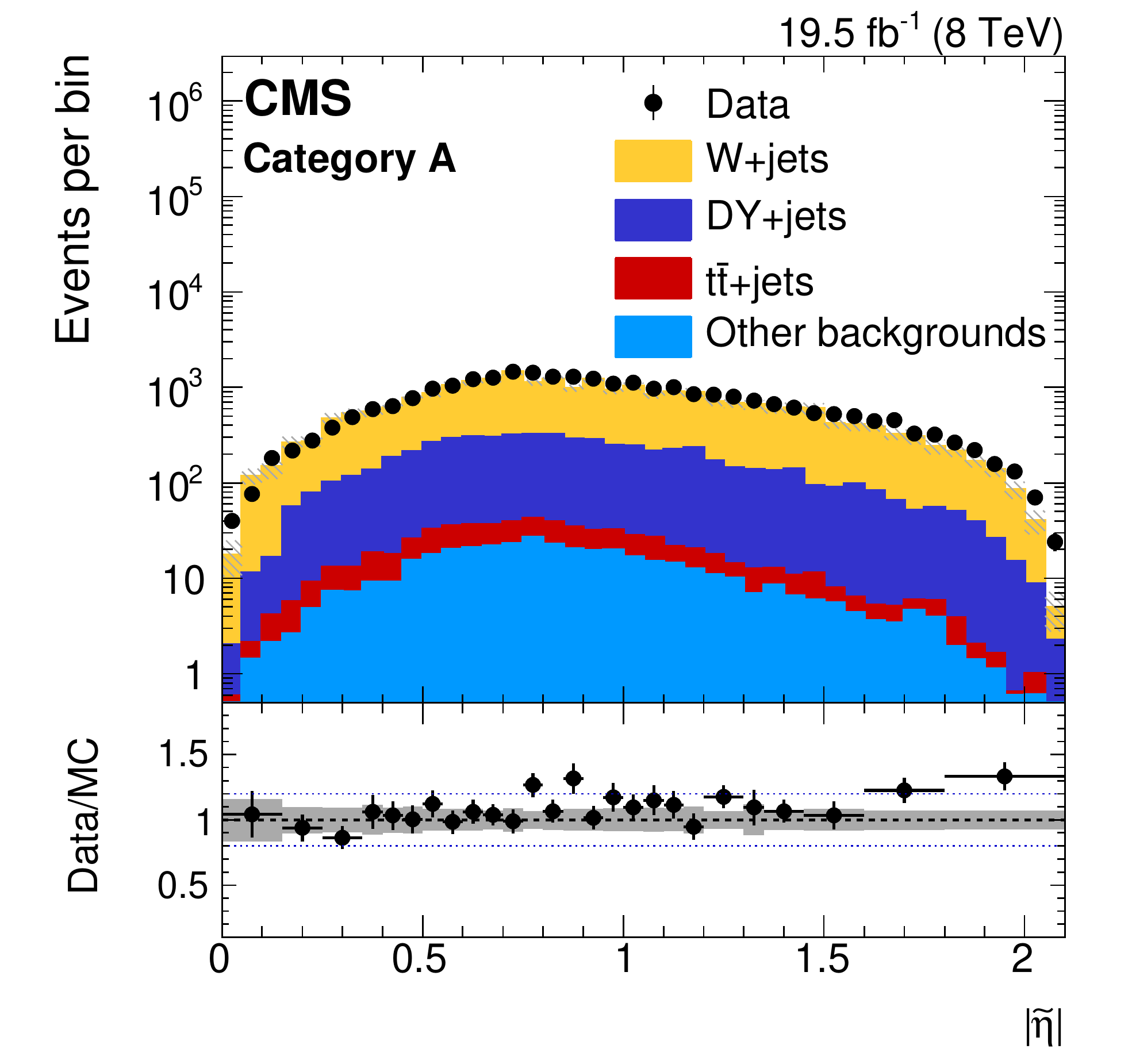}
\includegraphics[width=0.49\textwidth]{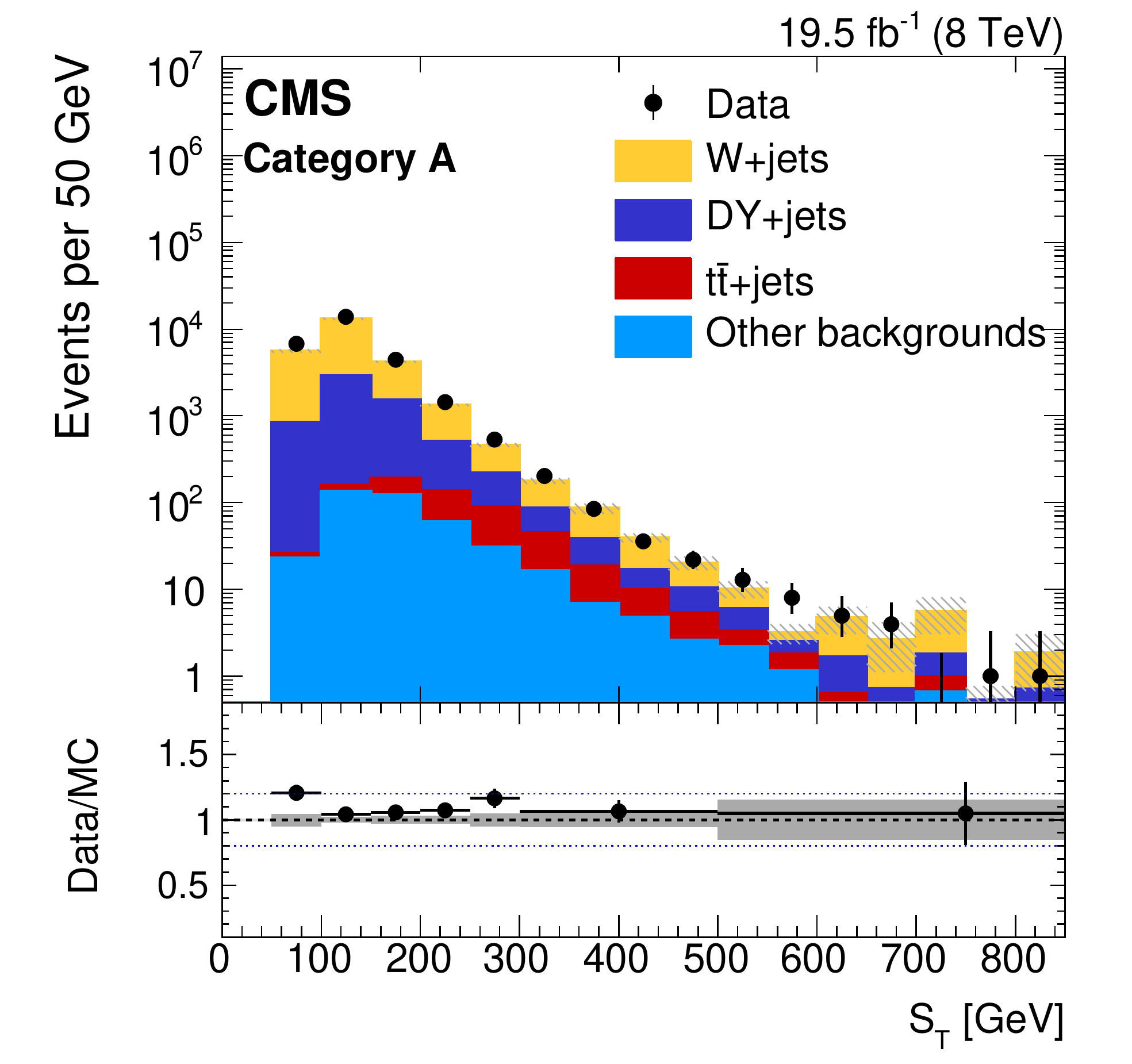}
\caption{Comparison between data and simulation in the $\widetilde{\abs{\eta}}$ (left) and $\ST$ (right) distributions using the signal-depleted selection of events in category A with a same-sign $\mu\tauh$ pair.
Other backgrounds refer to contributions predominantly from processes such as diboson and single top quark production, as well as QCD multijet and other rare SM processes detailed in Section~\ref{DataMCsamples}.
The hatched regions in the distributions and the shaded bands in the Data/MC ratio plots represent the statistical uncertainties in the expectations.
The data-simulation agreement is observed to be within 20\%, and is assigned as the normalization systematic uncertainty for the $\ttbar$+jets, DY+jets and diboson contributions in the signal region.
\label{PreselectionPlotsCatA}}
\end{figure}

A signal-depleted selection of events with a same-sign $\mu\tauh$ pair, created by vetoing events with more than one jet,
is used to check the performance and normalization of the simulated background samples.
In order to reduce the QCD multijet background contribution, an additional requirement of $M_\mathrm{T}(\mu,\ETmiss)>40$\GeV is imposed using the muon candidate in the selected same-sign $\mu\tauh$ pair.
Figure~\ref{PreselectionPlotsCatA} illustrates the agreement between data and simulation in the $\widetilde{\abs{\eta}}$ and $\ST$ distributions, which is found to be within 20\%.

\subsection{Event selection in category B \label{EventSelectionCatB}}

In category B, muon candidates are required to have $\pt>30$\GeV and $\abs{\eta}<2.1$.
The muon isolation requirement follows the tight muon selection defined for category A.

Hadronically decaying $\tau$ leptons must satisfy $\pt>20$\GeV and $\abs{\eta}<2.1$.
For $\tauh$ candidates a medium isolation requirement is used, where
the scalar $\pt$ sum of PF candidates must not exceed 1\GeV in a cone of radius $\Delta R=0.5$.
As in category A, $\tauh$ candidates must satisfy the requirement discriminating against misreconstructed electrons and muons.

Electron candidates are required to have $\pt>35$\GeV and $\abs{\eta}<2.1$.
The electron isolation requirement follows the description in category A, but with a tighter threshold at 10\% in order to match the trigger isolation requirements in the $\Pe\tauh$ channel.

Jets are required to have $\pt>30$\GeV, $\abs{\eta}<2.5$, and those overlapping with $\tauh$ candidates within a cone of $\Delta R=0.5$ are ignored.
Furthermore, $\tau$ leptons that overlap with a muon, and electrons that overlap with a jet within a cone of $\Delta R=0.5$ are not considered.

Each event is required to have at least one electron or muon and one $\tauh$ candidate.
Events containing muons are vetoed in the $\Pe\tauh$ channel.
Events satisfying the category A selection criteria are also vetoed, thus in the case of the $\mu\tauh$ selection, category B mostly consists of events with opposite-sign $\mu\tauh$ pairs.

In addition, events are required to have $\ST>1000$\GeV, $\ETmiss>50$\GeV, and at least three jets, where the leading and subleading jets further satisfy $\pt>100$ and 50\GeV, respectively.
The analysis in category B is performed in four search regions defined as a function of the transverse momentum of the leading $\tauh$ candidate: $20<\pt^\tau<60$\GeV, $60<\pt^\tau<120$\GeV, $120<\pt^\tau<200$\GeV, and $\pt^\tau>200$\GeV.
Since events with $\Pe\tauh$ and $\mu\tauh$ pairs are separated, this selection leads to eight search regions.
The two low-$\pt^\tau$ regions are mainly used to constrain the SM background processes, whereas the signal is expected to populate the two high-$\pt^\tau$ regions.
The selections on $\ST$, the momenta of the three jets, and $\ETmiss$ have been optimized with respect to the expected limits on the signal cross section obtained in the statistical evaluation of the search regions as described in Section~\ref{ResultsSection}.

\begin{figure}[htp]
\centering
\includegraphics[width=0.49\textwidth]{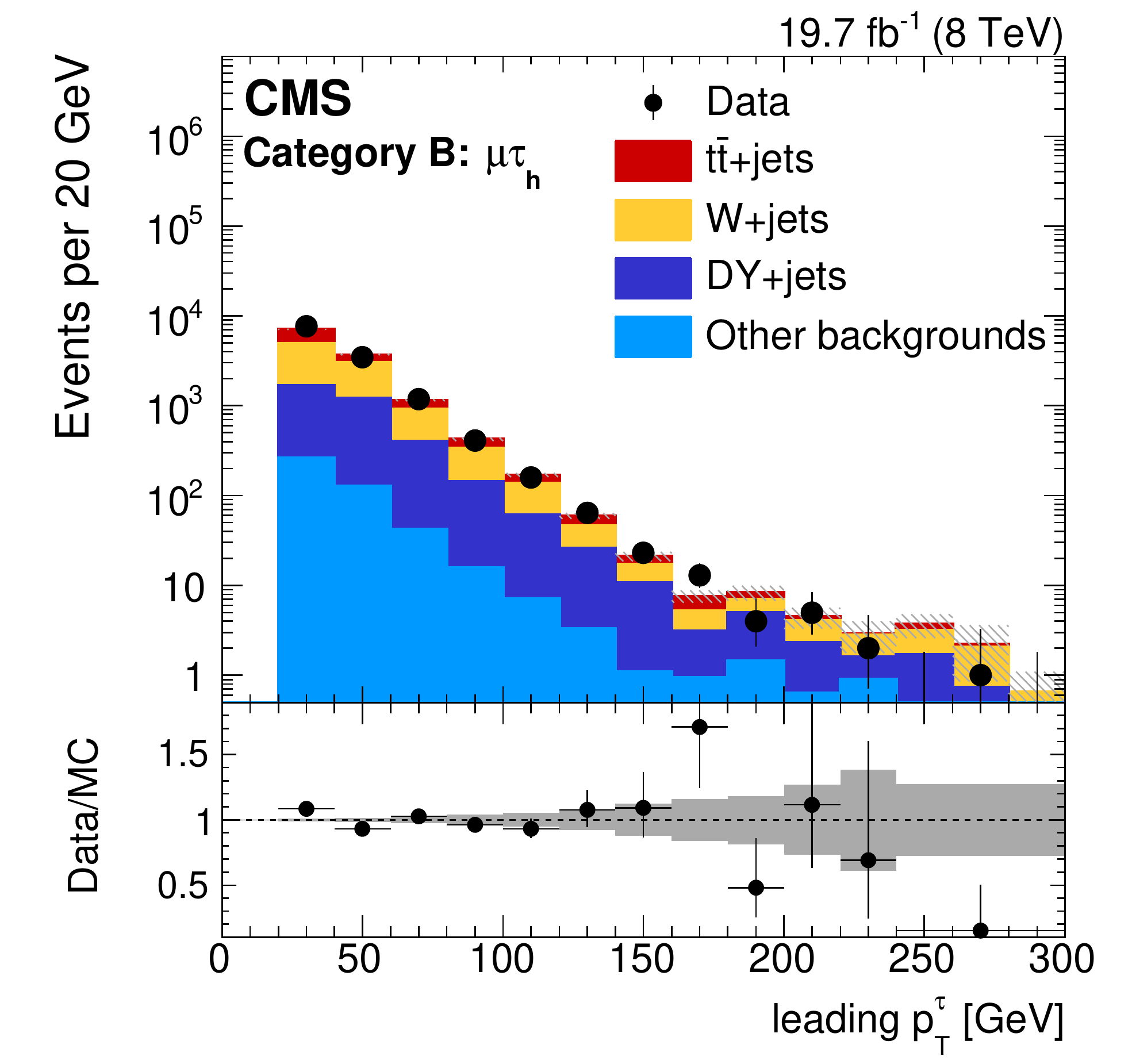}
\includegraphics[width=0.49\textwidth]{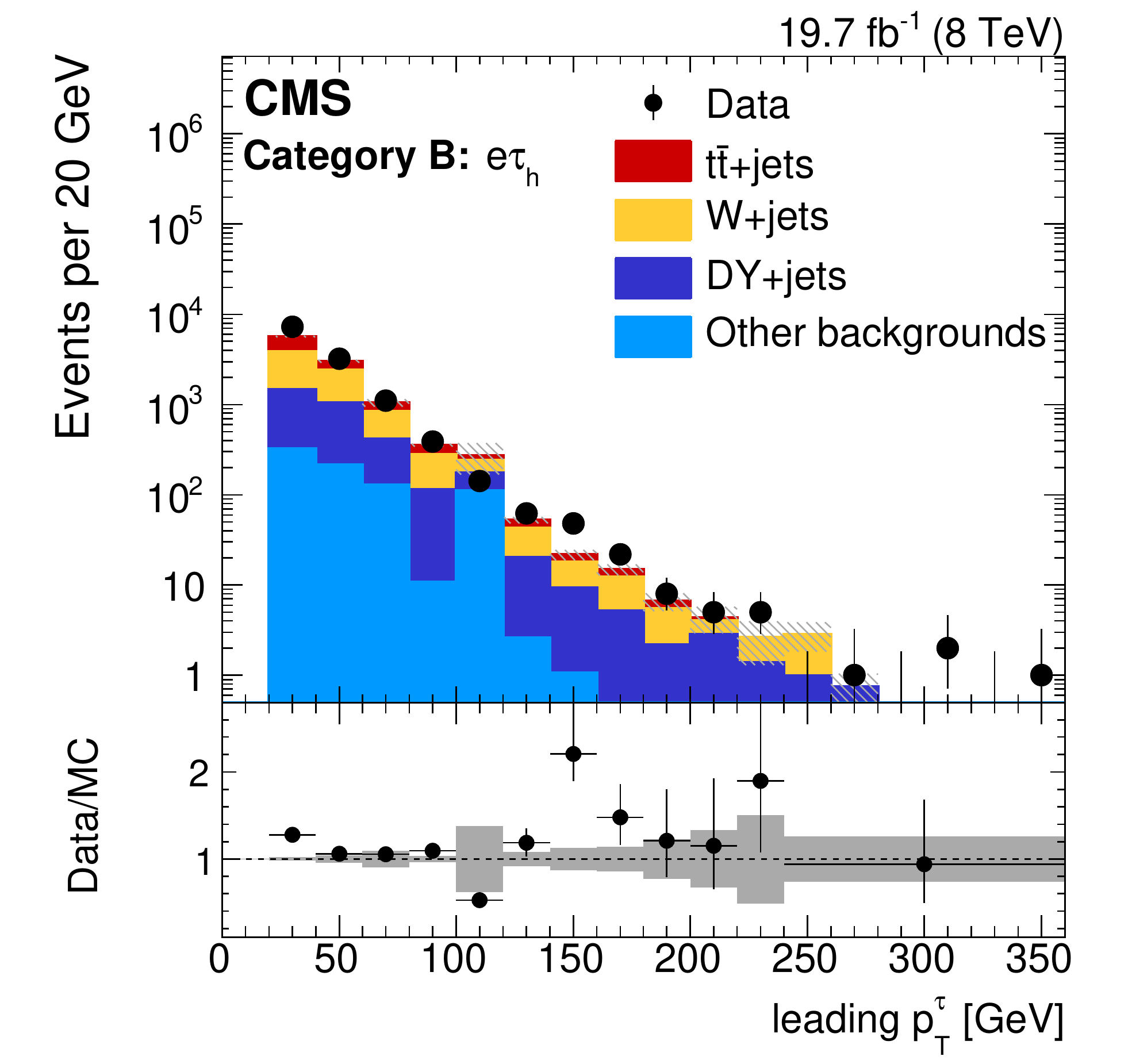}
\caption{Comparison between data and simulation in the leading $\tau$ lepton $\pt$ distributions using the signal-depleted selection of events in category B in the $\mu\tauh$ channel (left) and in the $\Pe\tauh$ channel (right).
Other backgrounds refer to contributions predominantly from processes such as diboson and single top quark production, but also include QCD multijet and rare SM processes detailed in Section~\ref{DataMCsamples}.
The hatched regions in the distributions and the shaded bands in the Data/MC ratio plots represent the statistical uncertainties in the expectations.
\label{PreselectionPlotsCatB}}
\end{figure}

A signal-depleted selection is used to check the performance of the simulated background samples in category B.
In this selection, events with $\ell\tauh$ pairs are required to have $\ETmiss < 50$\GeV and at least two jets with $\pt>50$\GeV, $\abs{\eta}<2.5$.
Figure ~\ref{PreselectionPlotsCatB} shows that in general the agreement between data and simulation is within the statistical uncertainties in the leading $\pt^{\tau}$ distributions.
In the $\Pe\tauh$ channel, a small excess in the  $\pt$ distribution is observed around 150\GeV.
As the other kinematic distributions in the signal-depleted region show no other significant deviations, the excess is assumed to be a statistical fluctuation.

\section{Backgrounds}

For this analysis, prompt leptons are defined to be those that come from the decays of W bosons, Z bosons or $\tau$ leptons, and are usually well isolated.
Leptons originating from semileptonic heavy-flavor decays within jets and jets misreconstructed as leptons are both labelled as misidentified leptons, and generally are not isolated.
In category A, the expected same-sign background events are mostly due to misidentified leptons, while category B has significant additional prompt-prompt contributions.
In accordance with the expected background compositions, data events are used to estimate the dominant misidentified lepton backgrounds in category A, eliminating the need to evaluate the simulation based systematic uncertainties, whereas the prompt-prompt backgrounds in category B require the consideration of these uncertainties.
Simulated samples corrected for $\tau$ lepton misidentification rates are used for the estimation of the backgrounds in category B.

\subsection{Backgrounds in category A\label{BackgroundsCatA}}

The same-sign dilepton requirement yields a background which mainly consists of events that contain misidentified leptons (especially jets misidentified as $\tau$ leptons).
These events come from semileptonic $\ttbar$+jets and W+jets processes in approximately equal proportions.
Smaller background contributions result from SM processes with genuine same-sign dileptons, such as diboson, $\ttbar\PW$, $\ttbar\Z$, and $\PW^\pm\PW^\pm\PQq\PQq$ events, and opposite-sign dilepton events in which the $\tauh$ charge has been misidentified,
such as DY+jets and fully leptonic $\ttbar$+jets events.
Events with misidentified leptons contribute up to 90\% of the total background, depending on the set of $\ST$ and $\tau$ lepton $\pt$ requirements, and are especially dominant in selections for $M_{\LQt}\le400$\GeV.

\subsubsection{Lepton misidentification}

Background contributions due to misidentified leptons are estimated using the observed data via a ``matrix method''~\cite{Chatrchyan:2012bra}.
For a given set of selection requirements, four combinations are defined based on the selection quality of the selected same-sign $\mu\tauh$ pair.
Events in which both leptons satisfy the tight selection requirements are classified as TT events, whereas those with both leptons failing the tight selection while satisfying the loose selection requirements are classified as LL events.
Similarly, events with only the muon or the $\tauh$ candidate satisfying the tight selection and with the other lepton satisfying the loose selection but failing the tight selection requirements are labeled as TL or LT events, respectively, where the muon is denoted first in the labeling.

The probabilities with which prompt ($p$) and misidentified ($m$) muon and $\tauh$ candidates pass a tight selection, given that they satisfy a loose selection, are measured as a function of the lepton $\pt$ in regions of $\ST$, lepton $\abs{\eta}$, and $\Delta R(\mu,j)_{\text{min}}$ or $\Delta R(\tau,j)_{\text{min}}$.
The TT events constitute the search region, whereas TL, LT, and LL events, together with the prompt and misidentification probabilities, are used to calculate the misidentified lepton contributions to the signal region, $N_{\mathrm{TT}}^{\text{misID}}$, as given in Eqs.~(\ref{inverseMatrix}) and~(\ref{TotalFakeBackground}).
\begin{equation}\label{inverseMatrix}
\begin{pmatrix}
N_{\mathrm{MM}}\\
N_{\mathrm{MP}}\\
N_{\mathrm{PM}}\\
N_{\mathrm{PP}}\end{pmatrix}
=
\frac{1}{(p_\mu-m_\mu)(p_\tau-m_\tau)}
\begin{pmatrix}
~~~p_\mu{\cdot}p_\tau & -p_\mu{\cdot}\widehat{p_\tau}  &-\widehat{p_\mu}{\cdot}p_\tau & ~~~\widehat{p_\mu}{\cdot}\widehat{p_\tau} \\
-p_\mu{\cdot}m_\tau &  ~~~p_\mu{\cdot}\widehat{m_\tau}   & ~~~\widehat{p_\mu}{\cdot}m_\tau  & -\widehat{p_\mu}{\cdot}\widehat{m_\tau} \\
-m_\mu{\cdot}p_\tau & ~~~m_\mu{\cdot}\widehat{p_\tau}    & ~~~\widehat{m_\mu}{\cdot}p_\tau  & -\widehat{m_\mu}{\cdot}\widehat{p_\tau} \\
~\;~~m_\mu{\cdot}m_\tau   & -m_\mu{\cdot}\widehat{m_\tau}    &-\widehat{m_\mu}{\cdot}m_\tau   & ~~~\widehat{m_\mu}{\cdot}\widehat{m_\tau}\end{pmatrix}
\begin{pmatrix}
N_{\mathrm{LL}} \\
N_{\mathrm{LT}} \\
N_{\mathrm{TL}} \\
N_{\mathrm{TT}} \end{pmatrix},
\end{equation}
\begin{equation}\label{TotalFakeBackground}
N_{\mathrm{TT}}^{\text{misID}}=m_\mu m_\tau  N_{\mathrm{MM}}+m_\mu p_\tau  N_{\mathrm{MP}}+p_\mu m_\tau N_{\mathrm{PM}}.
\end{equation}

$N$ denotes the number of events in a given combination, and MM, MP, PM, and PP labels denote the double-misidentified, muon misidentified, $\tauh$ misidentified, and double-prompt combinations, respectively.
The complementary prompt probability is given as $\widehat{p}=1-p$, and the complementary misidentification probability is given as $\widehat{m}=1-m$.

Muon and $\tau$ lepton prompt probabilities are measured in DY+jets enhanced data regions with $\Z\to \mu\mu$ and $\Z \to \tau\tau \to \mu\tauh$ decays, respectively, and in simulated $\ttbar$+jets,  W+jets and $\LQt$ events.
For the $\tau$ lepton misidentification probability measurements, a $\PW(\to \mu\nu)$+jets enriched data set with additional $\tauh$ candidates is used.
A QCD multijet enhanced data set with a single muon candidate is used for the muon misidentification probability measurements.
In simulated samples, the $\tau$ lepton misidentification probability measurement is conducted in W+jets, $\ttbar$+jets, and $\LQt$ samples, while the muon misidentification probability measurement is made in QCD multijet, $\ttbar$+jets, and $\LQt$ samples.

The individual prompt and misidentification probability measurements conducted using simulated samples are combined into a single value for each of the $p$ and $m$ bins.
For each of these, an average value and an associated uncertainty is calculated to account for the process dependent variations.
These simulation based values are then combined with correction factors derived from the $p$ and $m$ measurements in data,
to account for any bias in the simulated detector geometry and response, providing the values used in Eqs.~(\ref{inverseMatrix}) and (\ref{TotalFakeBackground}).
The resultant muon prompt probabilities vary from $(70\pm3)\%$ to $(95\pm3)\%$ for low and high \pt muons, whereas $\tau$ lepton prompt probabilities are around $(60\pm6)\%$.
The muon and $\tau$ lepton misidentification probabilities are measured to be about $(1\pm1)\%$ and $(14\pm2)\%$, respectively.

The matrix method yields consistent results for the misidentification backgrounds when applied to a signal-depleted selection of events in data and to simulated events in the signal region.
The expected yields are in agreement with the observations within 1.5 standard deviations in both selections.

\subsubsection{Charge misidentification and irreducible backgrounds}

The background contributions due to lepton charge misidentification and irreducible processes with same-sign $\mu\tauh$ pairs are estimated directly from the simulated samples.
These prompt-prompt contributions are calculated by requiring a match ($\Delta R<0.15$) between the reconstructed lepton candidate and a generator-level object of the same kind without any requirement on the charge.
The charge misidentification backgrounds are dominated by $\tauh$ candidates,
and these backgrounds contribute to 2--3\% of the total expected backgrounds in selections for $M_{\LQt}\le400$\GeV, whereas are negligible in those for higher $\LQt$ masses.

\subsection{Backgrounds in category B \label{osbackground}}

In category B, major background processes are $\ttbar$+jets, W+jets, and DY+jets events.
Smaller contributions come from single top quark, diboson, $\ttbar\Z$, and QCD multijet events.
Contributions from prompt-prompt $\ell\tauh$ pairs are mainly expected in fully leptonic $\ttbar$+jets events, as well as DY+jets events with $\Z \to \tau\tau \to \ell\tauh$ decays and diboson events.
In all other processes, the $\tauh$ candidates are expected to originate from misidentified jets.
The misidentified electron and muon contributions have been found to be negligible after applying isolation and \ETmiss requirements.
The background estimation in category B is obtained from simulated samples with various corrections applied to account for differences between data and simulation in the reconstruction and identification of misidentified $\tau$ lepton candidates.

The $\tau$ lepton misidentification rate is defined as the probability for a misidentified $\tau$ lepton candidate originating from a jet to satisfy the final $\tau$ lepton identification criteria used in the analysis. The corresponding correction factor for the simulation is defined as the ratio of the data and the simulation-based rates.
The misidentification rates in data and simulation are measured in $\PW(\to \ell\nu)$+jets enriched events, containing at least one $\tau$ lepton candidate.
The $\tau$ lepton candidate is used as a misidentified probe, and the results are parametrized as a function of the $\tau$ lepton $\pt$.
Additional parametrizations, such as $\ST$, jet multiplicity, and $\Delta R(\tau,j)_{\text{min}}$, reveal no further deviations between the data and simulation. Thus a one-dimensional parametrization as a function of the $\tau$ lepton $\pt$ is used to describe any discrepancy between data and simulation. A small discrepancy is observed in the distribution of scale factors as a function of the $\tau$ lepton $\eta$ for $\abs{\eta}>1.5$.
An additional uncertainty is therefore assigned to the $\tau$ lepton scale factors for misidentified $\tau$ leptons in this $\eta$ region.

Measurements based on data are corrected by subtracting the small contributions due to prompt $\tau$ leptons, muons, and electrons which are misidentified as $\tau$ lepton candidates using the predictions from the simulated samples.
The systematic uncertainties in the correction factors are estimated by varying the cross sections of the dominant simulated processes within their uncertainties~\cite{Chatrchyan:2013faa}.

The resulting correction factors on the $\tau$ lepton misidentification rate are found to be in the range of 0.6--1.1 for the four $\tau$ lepton $\pt$ regions.
These weights are applied to each misidentified $\tau$ lepton candidate in all simulated background processes.

A jet originating from gluon emission has a smaller probability of being misidentified as a $\tauh$ candidate than those originating from quarks.
Quarks tend to produce incorrectly assigned $\tau$ lepton candidates with a like-sign charge.
Therefore, an additional systematic uncertainty is assigned to the correction factors based on the flavor composition of jets misidentified as $\tau$ leptons.
To determine this uncertainty, the measurement of the $\tau$ lepton misidentification rate is repeated for each of the charge combinations of the $\ell\tauh$ pair, $\tauh^\pm\ell^\pm$ and $\tauh^\pm\ell^\mp$.
Because of the different production modes of \PWp and \PWm bosons at the LHC, the four charge combinations have different quark and gluon compositions.
An estimate of the maximally allowed variance in the probability of each quark or gluon type to be misidentified as a $\tau$ lepton is obtained via the comparison of the misidentification rate measurements in the four channels.
The uncertainties in the misidentification rates are scaled according to the different expected flavor compositions in the signal and $\PW(\to \ell\nu)$+jets enriched regions used for the misidentification rate measurements.

\subsection{Systematic uncertainties}

In category A, the backgrounds due to misidentified leptons are derived from data and the associated systematic uncertainties are calculated by propagating the uncertainties in the muon and $\tau$ lepton prompt and misidentification probability measurements.
The uncertainties in the background rate of misidentified leptons lie in the range of 21--28\% in the central channel and 21--36\% in the forward channel.

In category B, the uncertainties in the correction factors on the misidentification rate of $\tau$ leptons vary from 23--38\% for the lower three $\tau$ lepton $\pt$ regions and up to 58--82\% for the highest $\pt$ region in the $\mu\tauh$ and  $\Pe\tauh$ channels.
These uncertainties are propagated to the estimate of the background of misidentified hadronically decaying $\tau$ leptons by varying the correction factors applied to the simulation within their uncertainties.

Since both the signal efficiencies and the prompt-prompt contributions to the background in category A and all the signal and background estimates in category B are determined using simulated events, the following sources of systematic uncertainty are considered.

Normalization uncertainties of 20\%  are applied for $\ttbar$+jets, DY+jets and diboson processes in category A as observed in the signal-depleted region presented in Fig.~\ref{PreselectionPlotsCatA}.
An uncertainty of 30\% is applied for other rare SM process as motivated by the theoretical uncertainties in the NLO cross sections for processes such as $\ttbar\PW$, $\ttbar\Z$~\cite{ttWZ,ttW}, and triboson~\cite{Alwall:2014hca} production.
For category B, these uncertainties in the MC normalization vary in a range between 15\% and 100\% according to previous measurements~\cite{Chatrchyan:2013faa}.
The CMS luminosity used in the normalization of signal and MC samples has an uncertainty of 2.6\% \cite{CMS-PAS-LUM-13-001}.

In order to account for uncertainties in the efficiency of $\tau$ lepton identification, an uncertainty of 6\% is applied for each prompt $\tau$ lepton found in the event.
The uncertainty in the $\tau$ lepton energy is taken into account by varying the energy of all $\tau$ leptons by $\pm$3\%.
Uncertainties induced by the energy resolution of prompt $\tau$ leptons in simulated samples are estimated by changing the resolution by $\pm$10\%.

Muon and electron identification, isolation, and trigger efficiencies are determined with a tag-and-probe method~\cite{TagAndProbe} in DY+jets enriched data.
In both categories, the muon reconstruction and isolation uncertainty is about 1\% and the single muon trigger matching uncertainty is $\leq$0.5\%.
Uncertainties in electron identification, isolation, and trigger efficiencies are considered only in the $\Pe\tauh$ channel of category B.
These uncertainties are \pt- and $\eta$-dependent and are found to be 0.3\% for electrons in the central detector region with $\pt<50$\GeV and up to 25\% for electrons with $\pt>500$\GeV.

Uncertainties in the jet energy resolution~\cite{JEScalib} are taken into account by changing the correction factors within their uncertainties.
These correction factors lie between 1.05 and 1.29 depending on jet $\eta$, with corresponding uncertainties varying from 5\% to 16\%.
The \pt- and $\eta$-dependent scale factors for the jet energy scale~\cite{JEScalib} are similarly varied by one standard deviation to obtain the corresponding uncertainties in simulated samples.
This corresponds to a 1--3\% variation of the scale factors.

The energy scale and resolution uncertainties in $\tau$ lepton, muon, electron, and jet candidates are also propagated in the calculation of $\ETmiss$ and $\ST$.

The uncertainty in the pileup re-weighting of simulated samples is estimated by varying the total inelastic cross section~\cite{Chatrchyan:2012nj} by 5\%.
Signal samples are produced with the CTEQ6L1 PDF set and the associated PDF uncertainties in the signal acceptance are estimated using the PDF uncertainty prescription for LHC~\cite{Alekhin:2011sk,PDF4LHC,Bourilkov}.
In category B, the PDF uncertainties are also calculated for the background processes estimated using simulations.

Additional uncertainties in major SM processes estimated from simulations are considered in category B.
Uncertainties in the factorization and normalization scales, $\mu_\mathrm{r}$ and $\mu_\mathrm{f}$, respectively, on $\ttbar$+jets and W+jets events are calculated by changing the corresponding scales by a factor of 2 or 0.5.
The effect of an uncertainty in the jet-parton matching threshold in the simulation of W+jets processes is evaluated by varying it within a factor of 2.
The uncertainty in the top quark \pt re-weighting procedure is estimated by doubling and removing the correction factors.

Table~\ref{SystematicsTable} shows a summary of the systematic uncertainties for categories A and B.

\begin{table}[htb]
\centering
\topcaption{Systematic uncertainty sources and their effects on background ($B$) and signal ($S$) estimates. 
Uncertainties affecting the signal yields in both categories and the background yields in category A are calculated using the selection criteria for the $M_{\LQt}=550$\GeV hypothesis. 
In category A, the uncertainties are reported for central/forward channels separately, where appropriate. 
In category B, all uncertainties are averaged over the four $\pt^\tau$ search bins. 
All values are symmetric except for the PDF uncertainty in the signal acceptance in category A, and the $\ttbar$ factorization and normalization scale uncertainty in category B. 
The $\tau$ misidentification rate uncertainties considered in category B are included in the matrix method uncertainty in category A. 
All uncertainties in the background estimates are scaled according to their relative contributions to the total expected background. 
}\label{SystematicsTable}
\resizebox{\linewidth}{!}{
\begin{tabular}{ l | c | c c | c c c c}
\hline
 &  &  \multicolumn{2}{c |}{Category A}  &  \multicolumn{4}{c}{Category B} \\ \cline{3-8}
 &  &  &  &  \multicolumn{2}{c}{$\mu\tauh$ ch.} &  \multicolumn{2}{c}{$\Pe\tauh$ ch.} \\
Systematic uncertainty & Magnitude (\%) & $B$ (\%) & $S$ (\%) & $B$ (\%) & $S$ (\%) & $B$ (\%) & $S$ (\%) \\
\hline
Integrated luminosity             & 2.6 & 0.4/1.2   & 2.6 & 2.6 & 2.6 & 2.6 &2.6  \\
Electron reco/ID/iso $\&$ trigger & $\pt$, $\eta$ dependent & --- & --- & --- & --- & 1.4 & 2.2 \\
Muon reco/ID/iso $\&$ trigger & 1.1 & 0.1/0.5  & 1.1  & 0.9 & 0.9 &  ---    &  ---    \\
$\tau$ lepton reco/ID/iso & 6.0 & 0.8/2.8   & 6.0     & 1.5 & 3.0 & 0.6 & 3.1 \\
Muon momentum scale $\&$ resolution & $\pt$ dependent & 0.1/0.3 & 0.4 & --- & --- & --- & --- \\
$\tau$ lepton energy scale       & 3.0  & 1.2/4.1   & 2.0 &  2.3 & 2.7 &0.6 & 1.5 \\
$\tau$ lepton energy resolution  & 10.0 & 0.2/0.8   & 0.9 & 1.2 & 1.3 & 0.2 & 0.1 \\
Jet energy scale       & $\pt$, $\eta$ dependent  & 0.9/3.2   & 1.9 & 4.2 & 1.9 &5.6 & 2.7  \\
Jet energy resolution  & $\eta$ dependent         & 0.4/1.2   & 1.0 & 0.8 & 0.3 &1.6 & 0.8\\
Pileup                 & 5.0 & 0.1/1.2   & 1.0/2.5 & 0.8 & 0.3 & 0.9 & 0.5 \\
PDF (on acceptance)           & ---   & ---        & $^{+2.9}_{-4.3}$$\big /$$^{+2.4}_{-6.2}$ & ---     & 0.7 & ---     & 0.9 \\
PDF (on background)           & ---   & ---        &                   ---                     & 8.7 & ---     & 8.3 & ---     \\
Matrix method  & --- & 23.1/15.3 & --- & --- & --- & --- & --- \\
$\text{Jet} \to \tau$ misidentification rate  & $\pt$ dependent & --- & --- & 8.2 & 1.0& 10.9 & 0.8 \\
$\Pe \to \tau$ misidentification rate &$\eta$ dependent & --- & --- &0.1& 0.1&0.1&   0.1\\
$\ttbar$ factorization/renormalization & $^{+100}_{-50}$  & --- & --- & $^{+6.1}_{-5.9}$ & --- &  $^{+2.9}_{-2.7}$ & --- \\
Top quark $\pt$ re-weighting  &  $\pt$ dependent  & --- & --- & 0.1 & --- & 0.1  & --- \\
W+jets factorization/renormalization & $^{+100}_{-50}$  & --- & --- &   4.3 & --- & 0.3 & --- \\
W+jets matching threshold & $^{+100}_{-50}$  & --- & --- & 1.3 & --- &  2.5 & --- \\
\hline
\end{tabular}}
\end{table}

\section{Results \label{ResultsSection}}

\begin{figure}[htbp]
\centering
\includegraphics[width=0.42\textwidth]{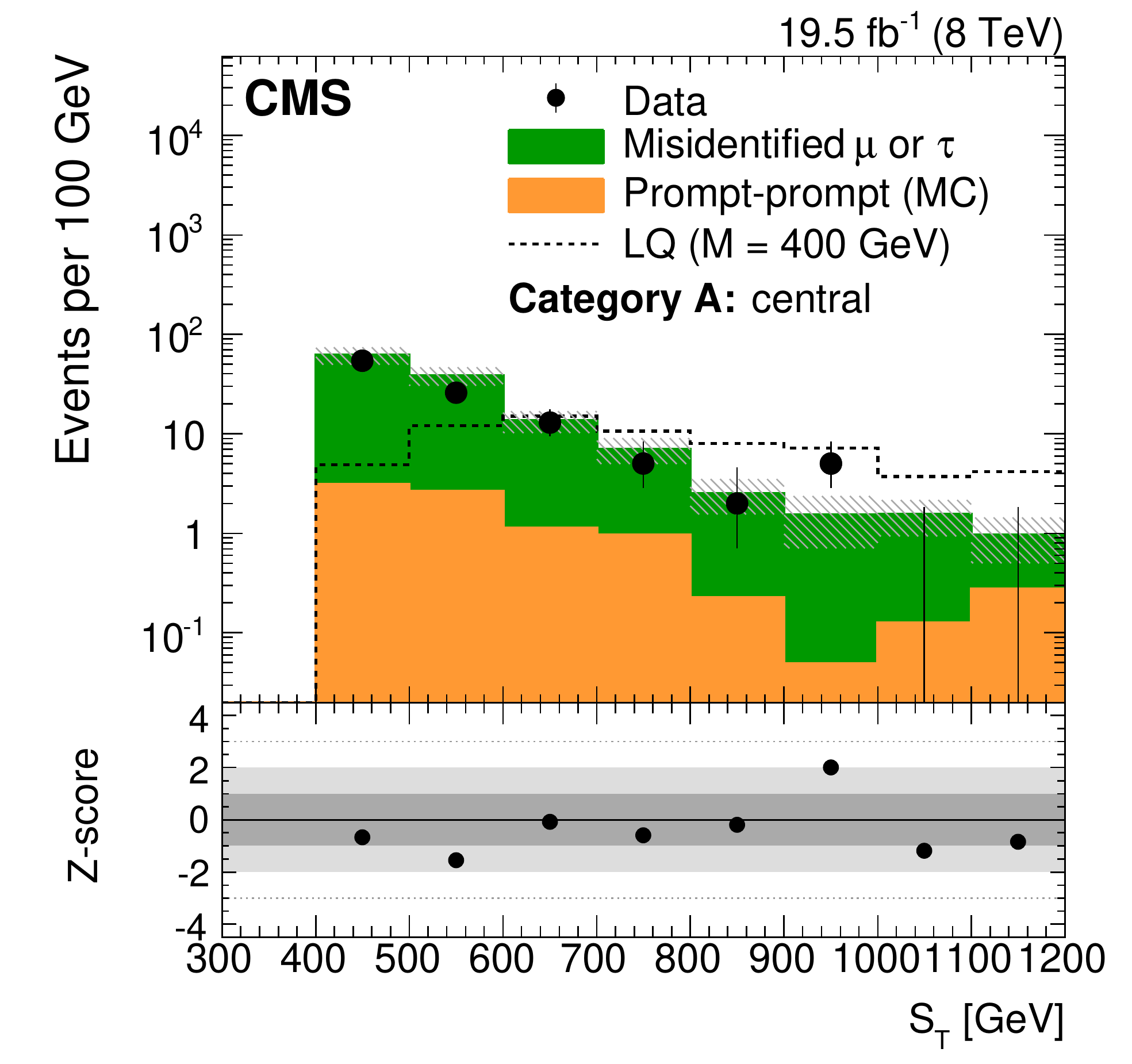}
\includegraphics[width=0.42\textwidth]{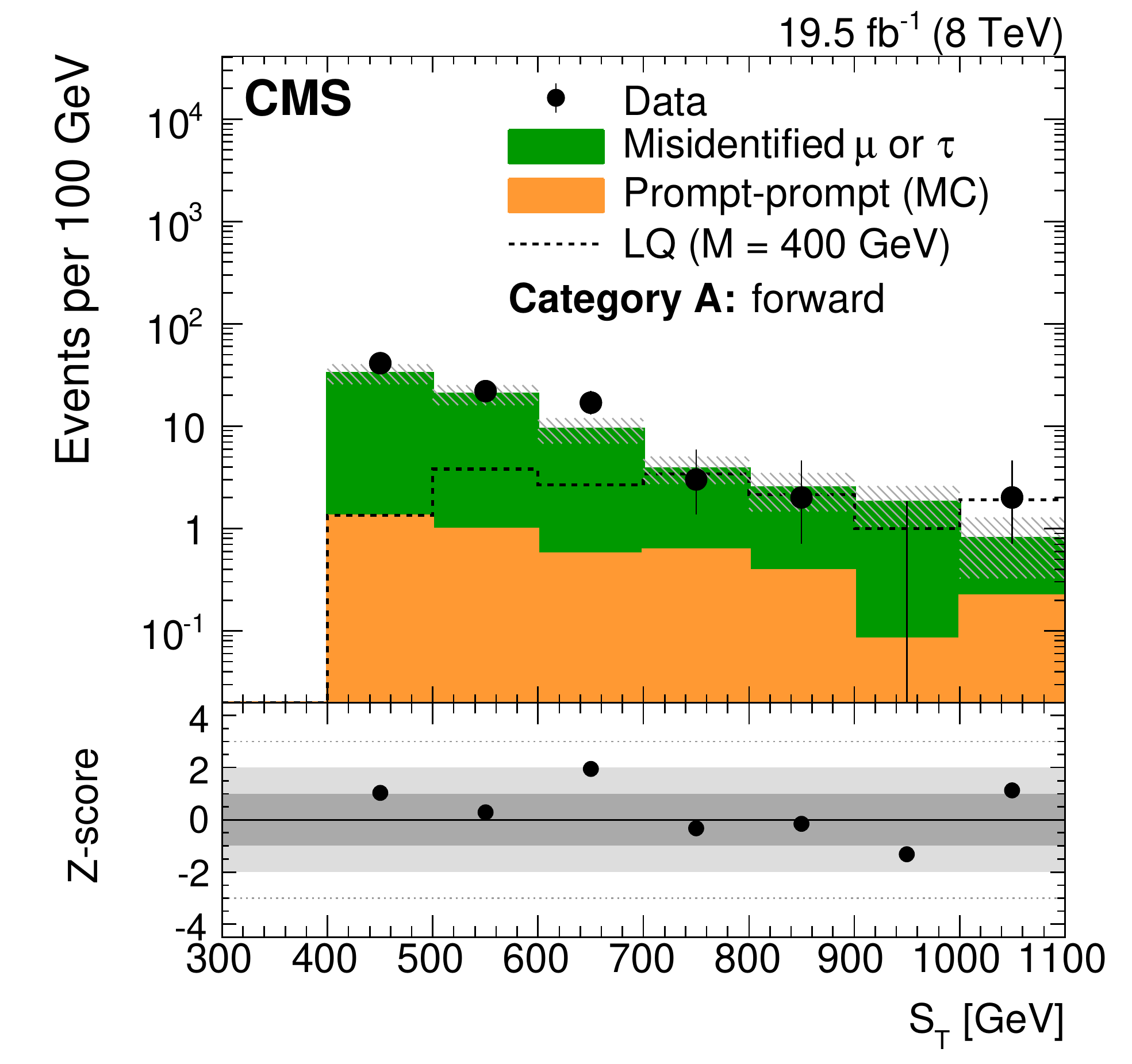}
\includegraphics[width=0.42\textwidth]{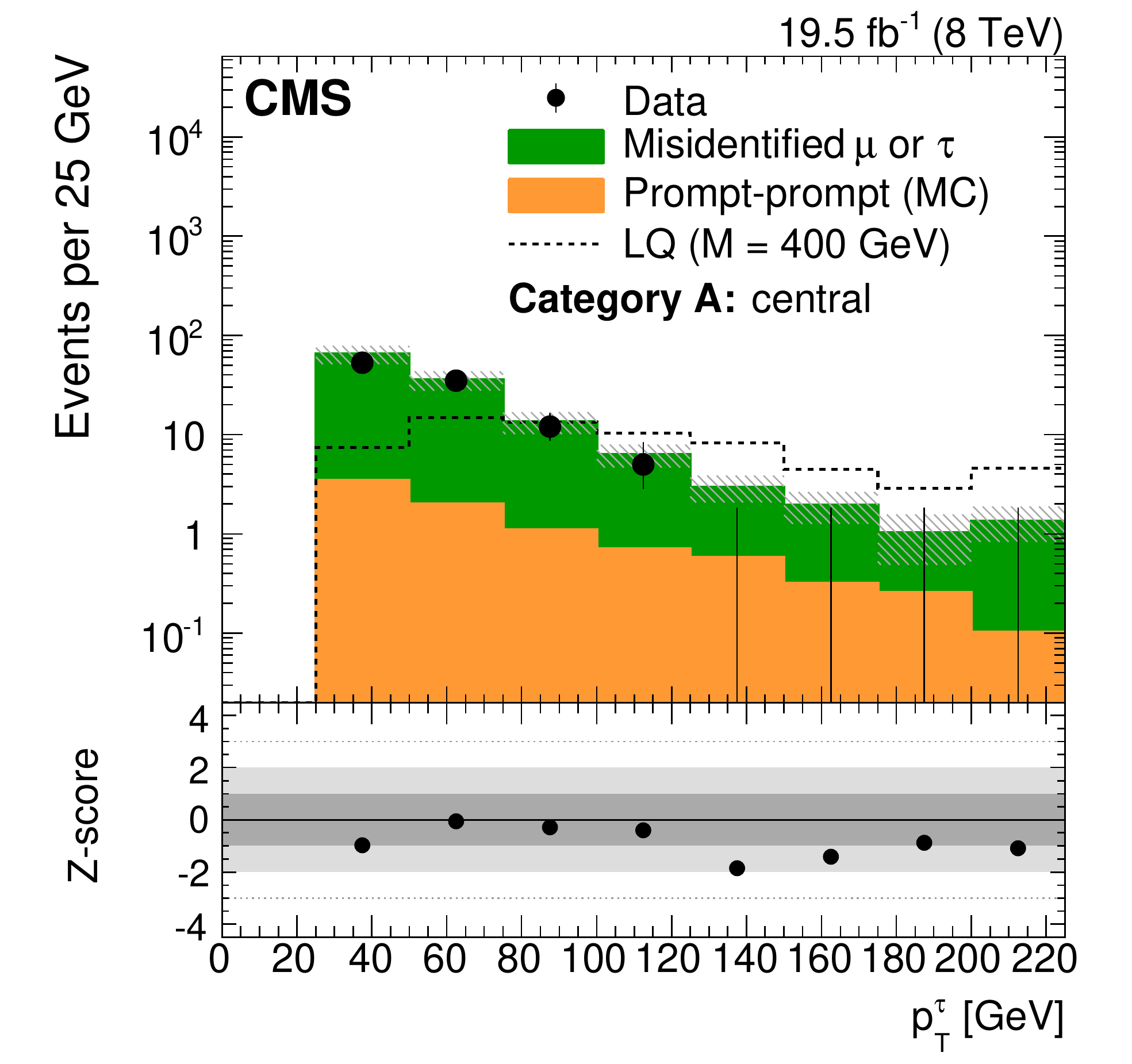}
\includegraphics[width=0.42\textwidth]{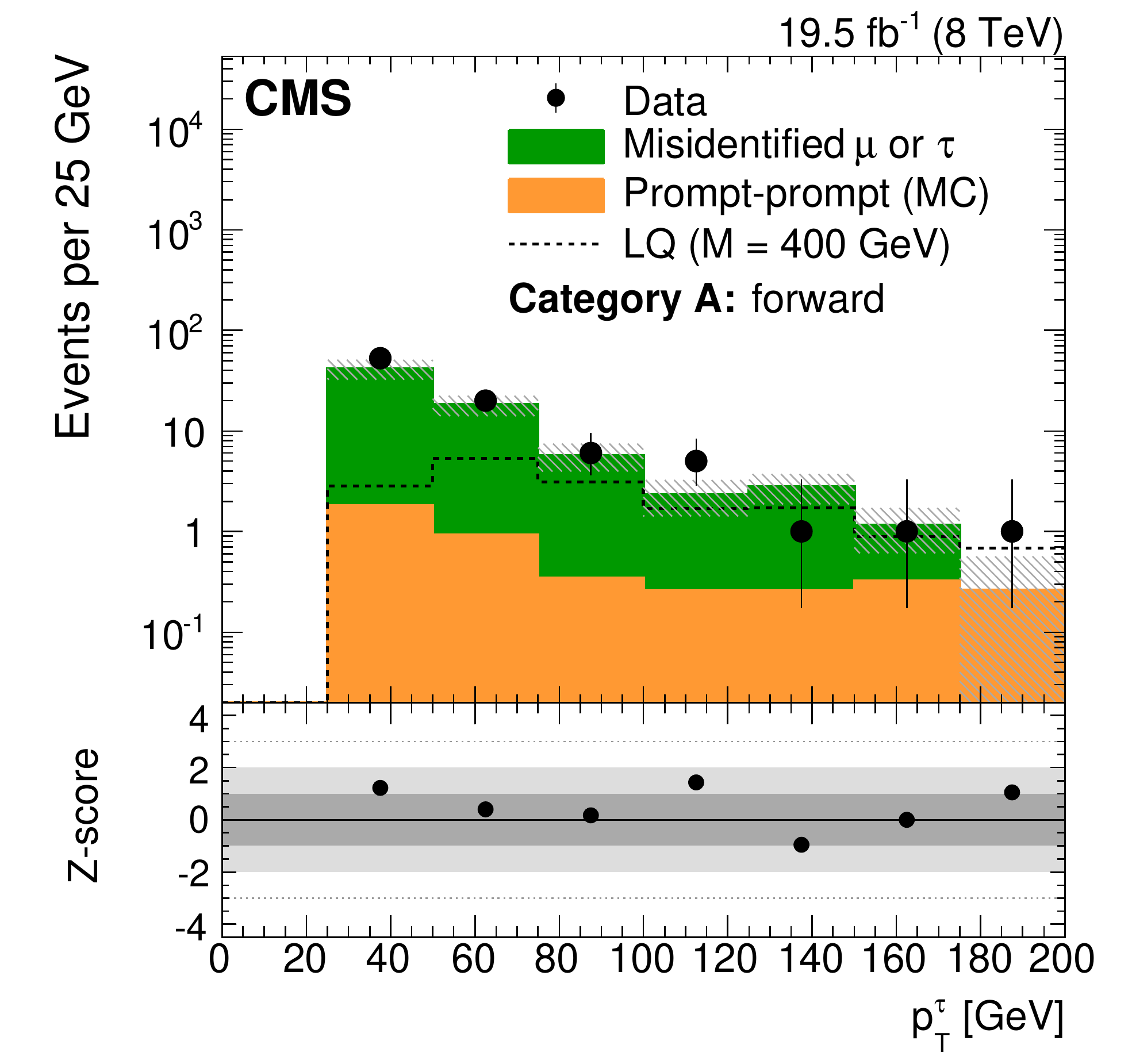}
\includegraphics[width=0.42\textwidth]{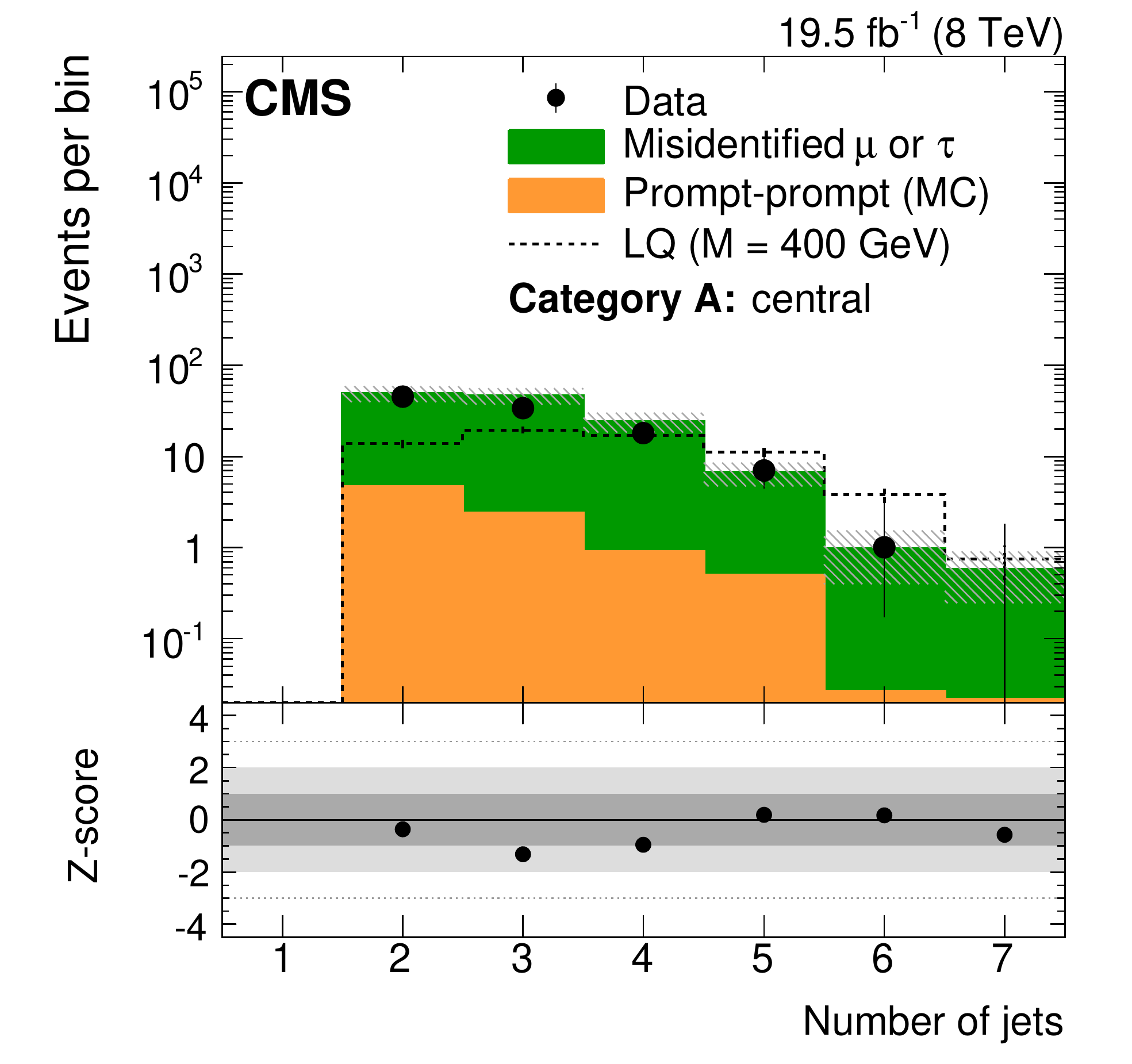}
\includegraphics[width=0.42\textwidth]{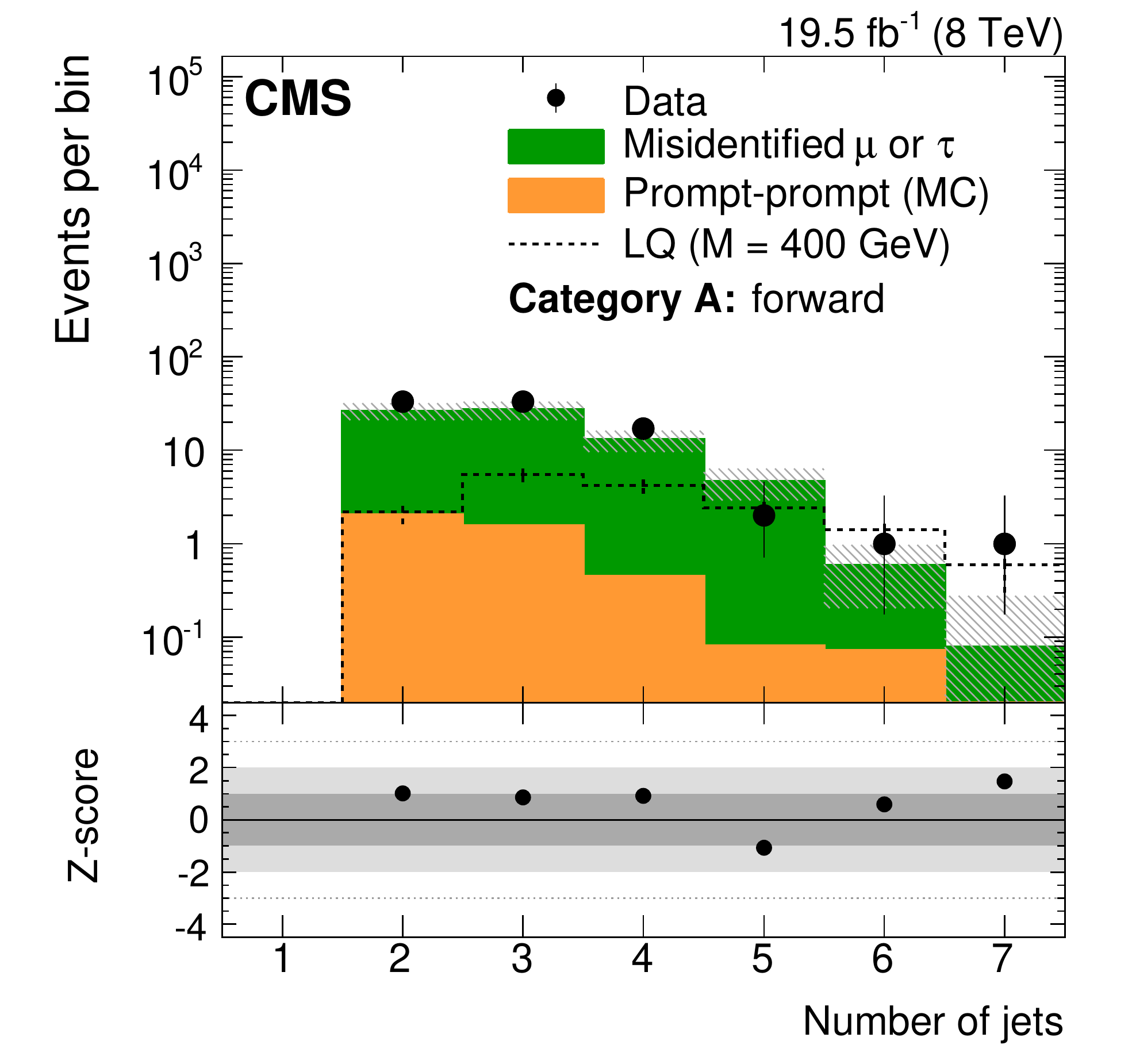}
\caption{The $\ST$, $\tau$ lepton $\pt$, and jet multiplicity distributions in the signal region of category A for central (left column) and forward (right column) channels, using the optimized selection for $M_{\LQt}=200$\GeV (all other optimized selection criteria yield events that are a subset of this selection).
The rightmost bin of each distribution includes overflow and no statistically significant excess is observed in the suppressed bins.
The systematic uncertainty for each bin of these distributions is determined independently.
Shaded regions in the histograms represent the total statistical and systematic uncertainty in the background expectation.
The Z-score distribution is provided at the bottom of each plot.
\label{DATASignalRegionPlots}}
\end{figure}

\begin{figure}[htbp]
\centering
\includegraphics[width=0.42\textwidth]{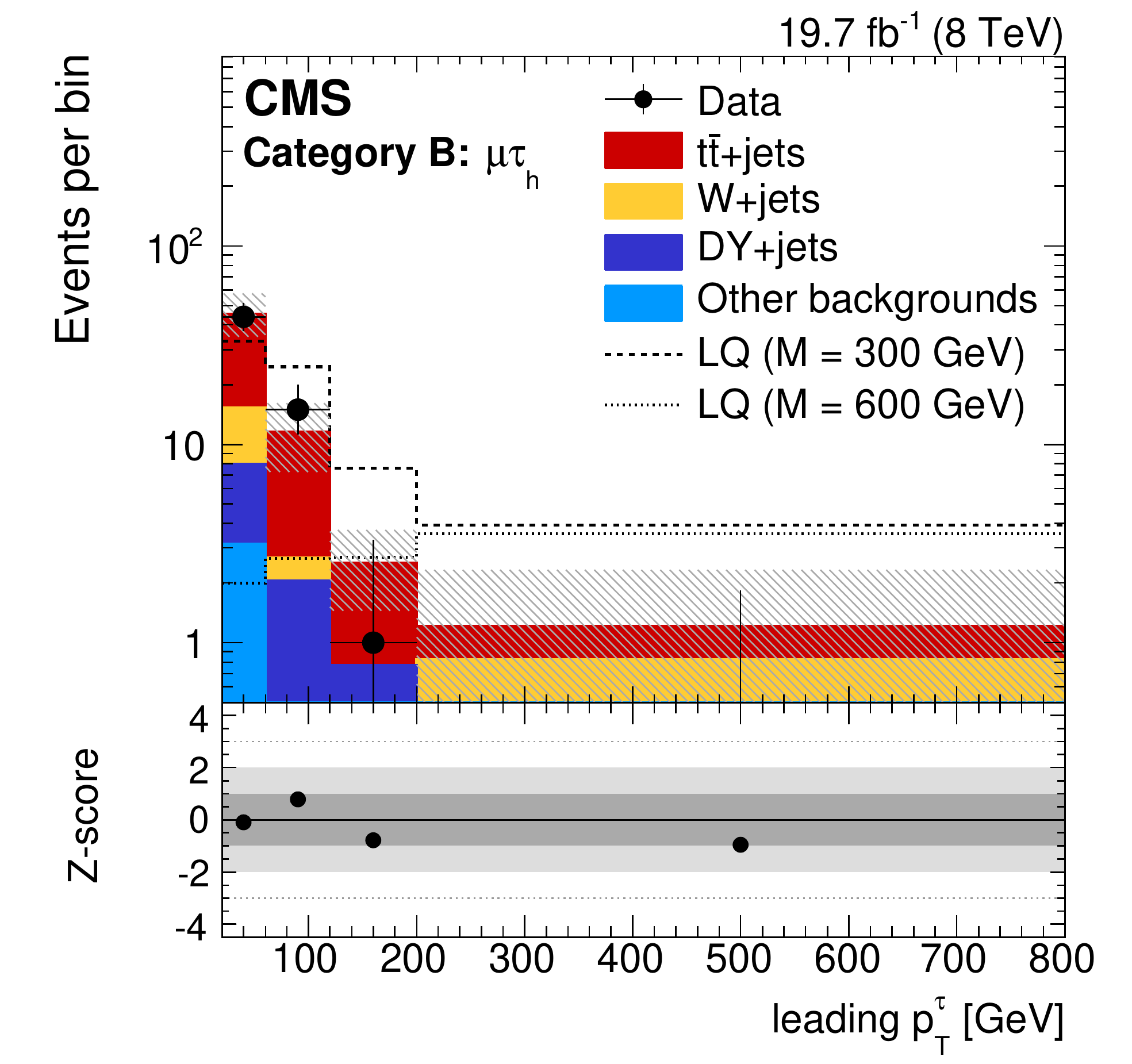}
\includegraphics[width=0.42\textwidth]{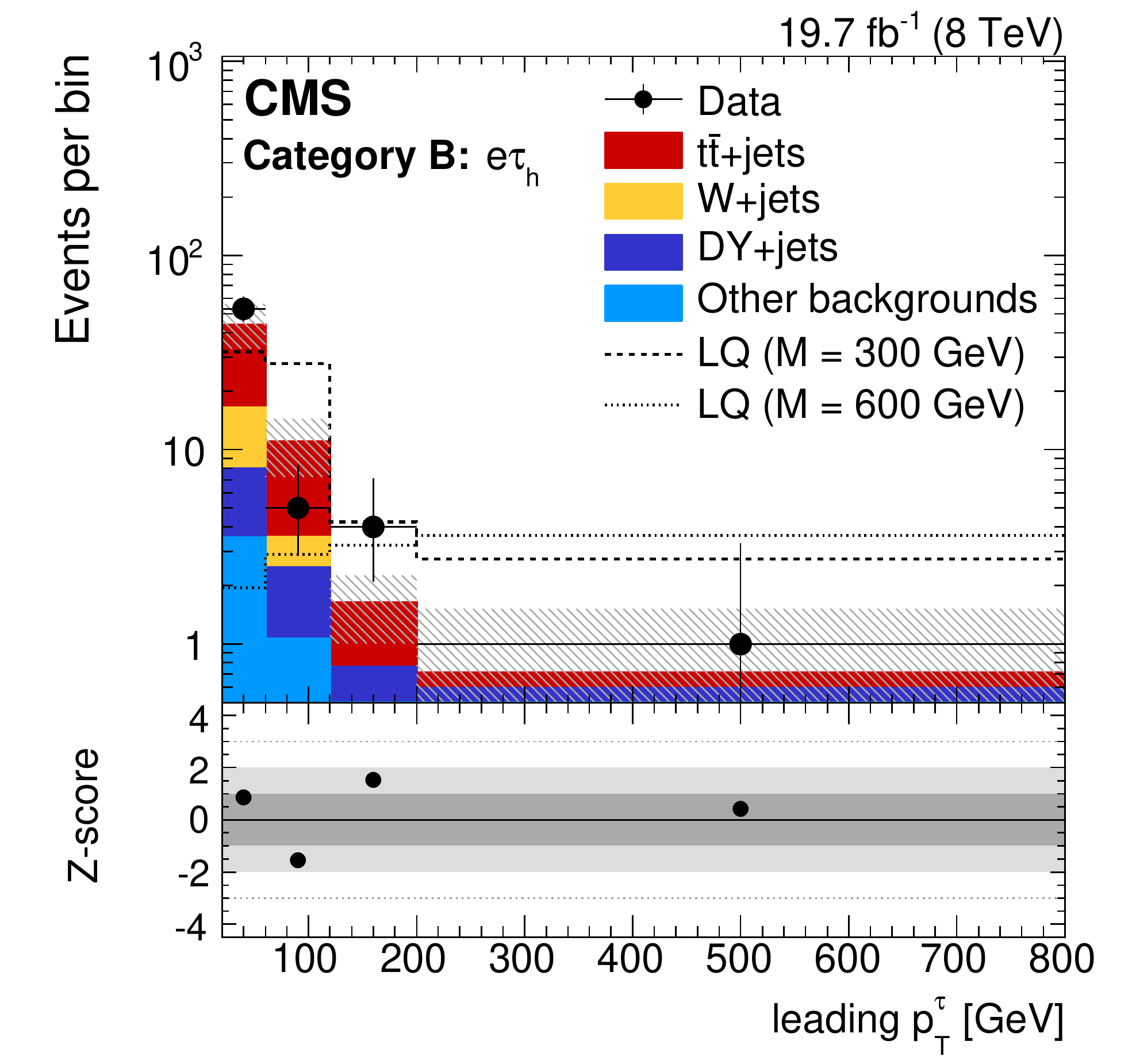}
\includegraphics[width=0.42\textwidth]{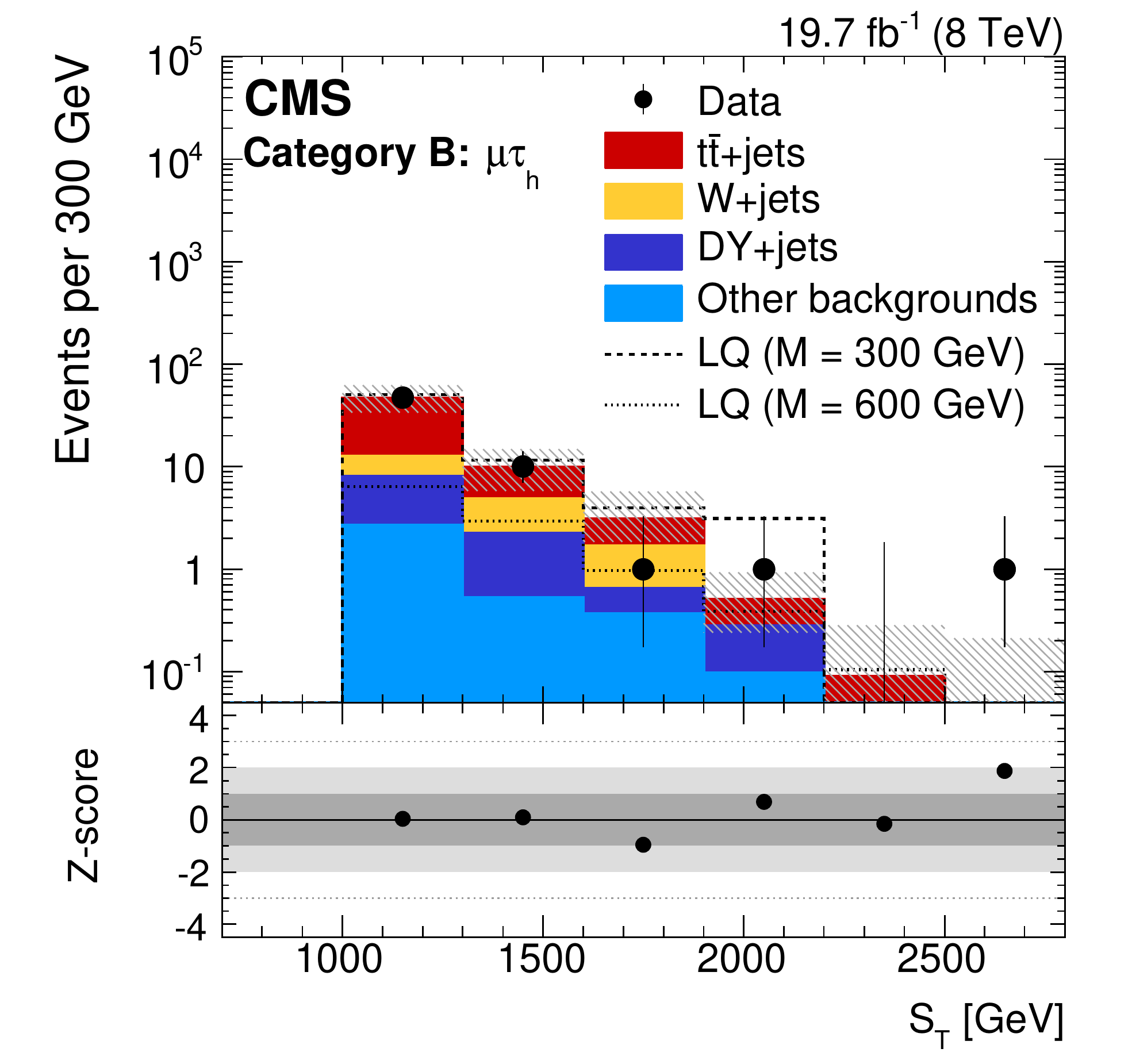}
\includegraphics[width=0.42\textwidth]{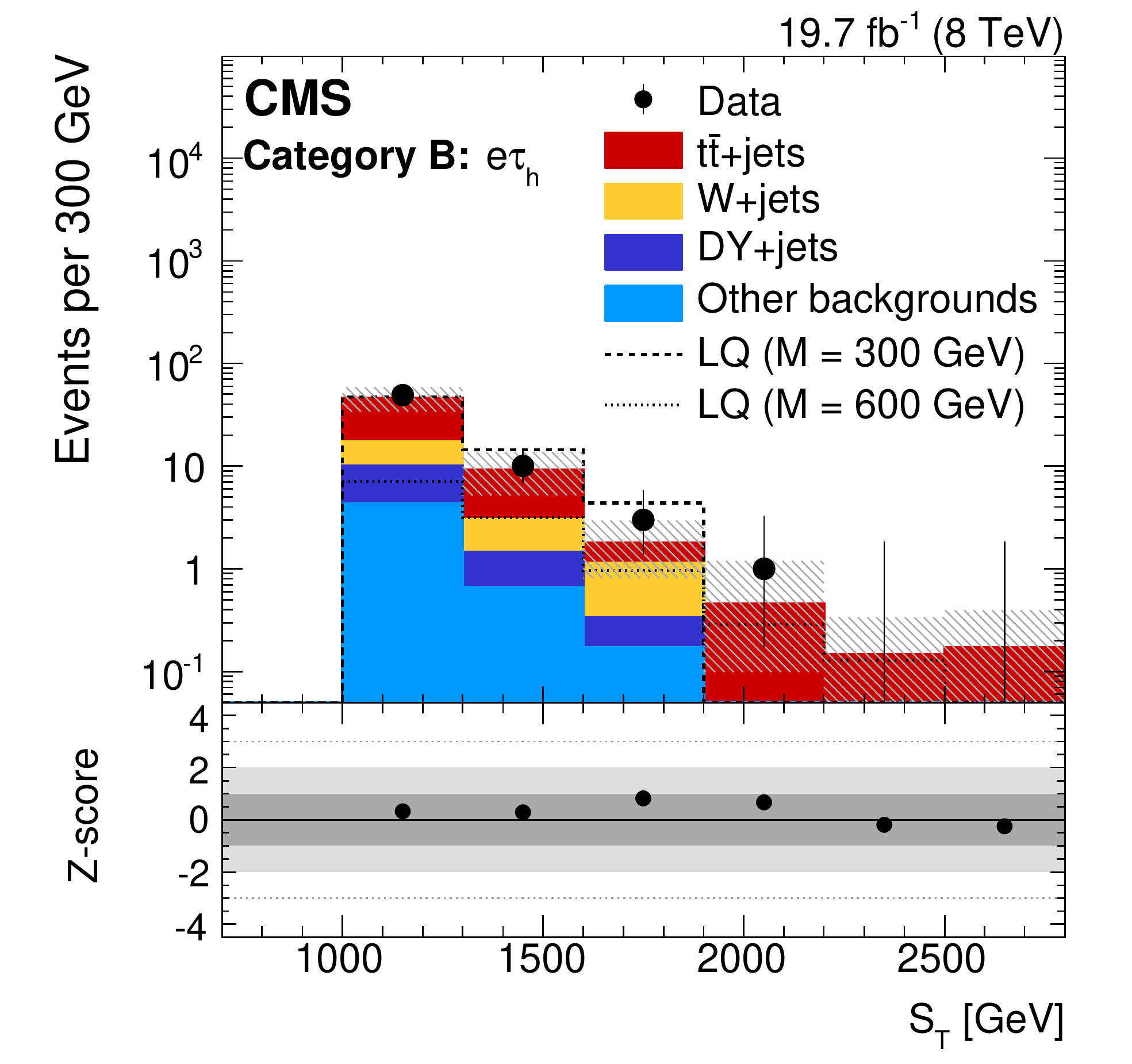}
\includegraphics[width=0.42\textwidth]{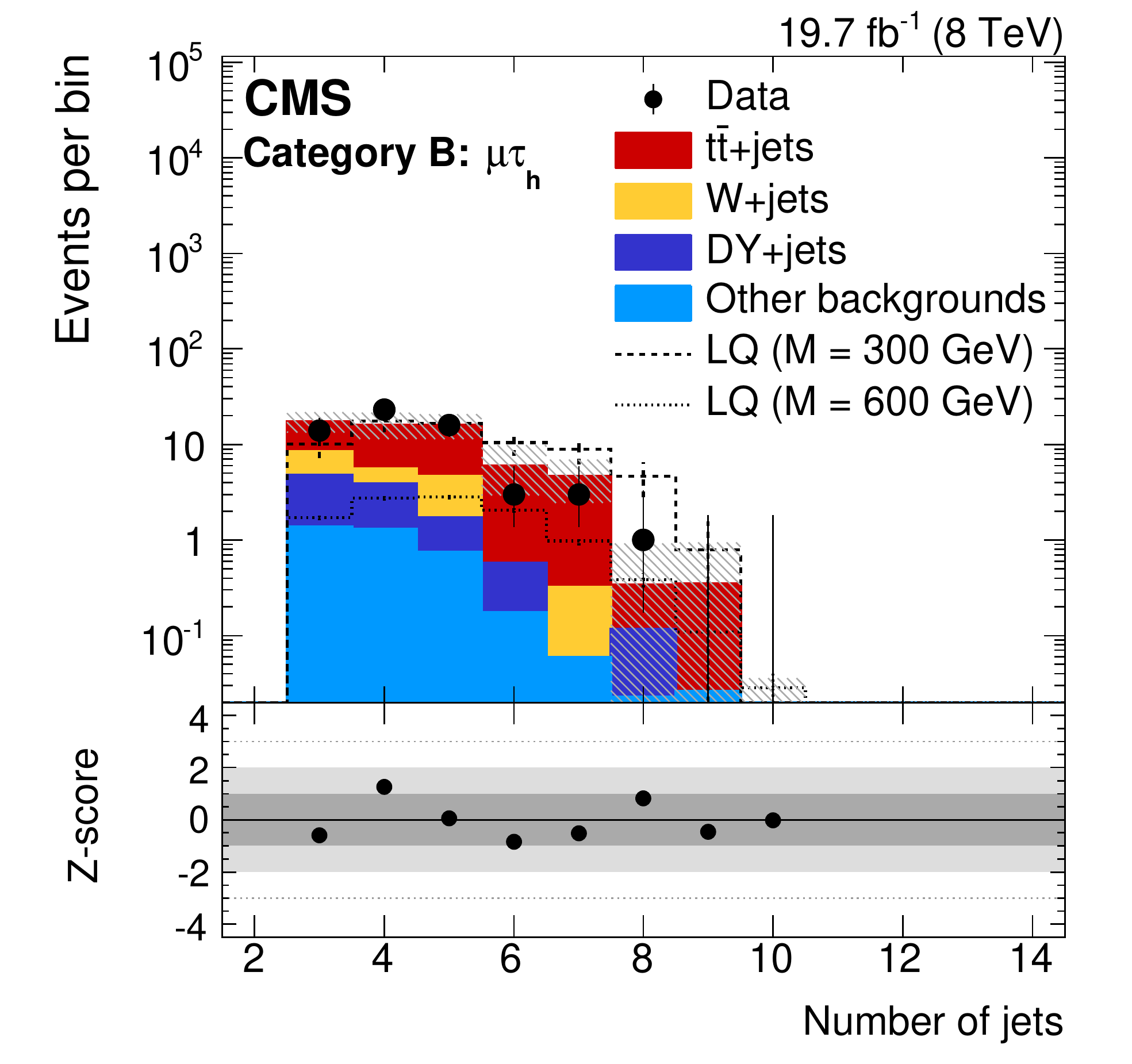}
\includegraphics[width=0.42\textwidth]{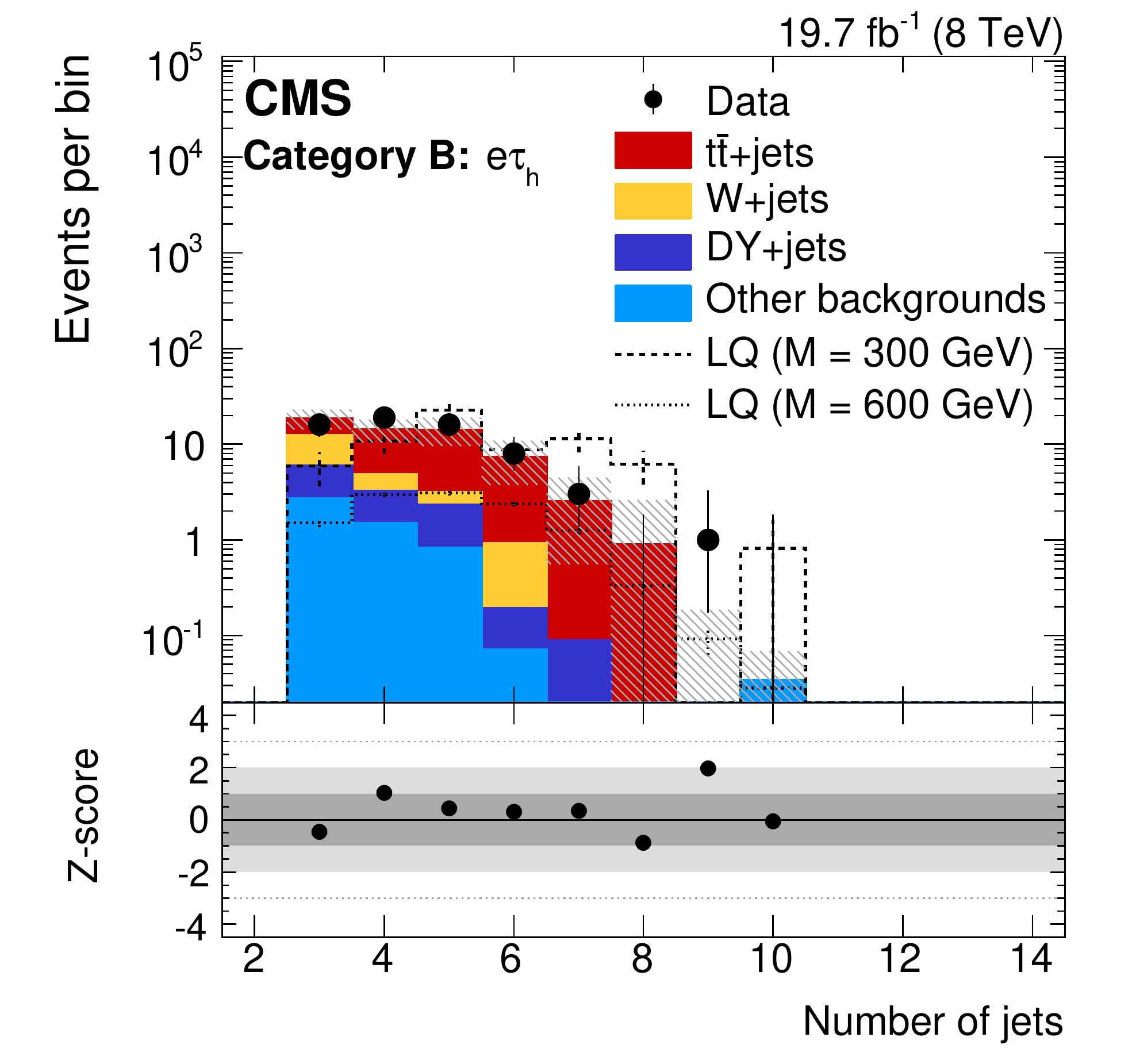}
{\caption{The leading $\tau$ lepton $\pt$, $\ST$, and jet multiplicity distributions in the signal region of category B for $\mu\tauh$ (left column) and $\Pe\tauh$ (right column) channels.
The rightmost bin of each distribution includes overflow and no statistically significant excess is observed in the suppressed bins.
Shaded regions in the histograms represent the total statistical and systematic uncertainty in the background expectation.
The Z-score distribution is provided at the bottom of each plot.
The four regions of the $\tau$ lepton $\pt$ correspond to the four search regions.}\label{DATASignalRegionPlotsOS}}
\end{figure}

The search results for category A (B) are presented in Table~\ref{SignalRegionResultsTable} (\ref{SignalRegionResultsTableOS} and \ref{SignalRegionResultsTableOSele}).
Figures~\ref{DATASignalRegionPlots} and~\ref{DATASignalRegionPlotsOS} show the comparison of data and the predicted backgrounds as a function of $\ST$, $\tau$ lepton $\pt$, and jet multiplicity parameters.
The dashed curves show the expectation for LQ signals.
For the comparison of expected and observed number of events in Tables \ref{SignalRegionResultsTable}--\ref{SignalRegionResultsTableOSele} and Figs.~\ref{DATASignalRegionPlots}--\ref{DATASignalRegionPlotsOS}, Z-scores are used. These are computed taking into account the total uncertainty in the mean number of expected events. A unit Z-score, $\abs{Z}=1$, refers to a two-tailed 1-standard deviation quantile ($\sim$68\%) of the normal distribution.
For each selection, the observed number of events is found to be in an overall agreement with the SM-only hypothesis and the distributions reveal no statistically significant deviations from the SM expectations.

A limit is set on the pair production cross section of charge $-1/3$ third-generation scalar LQs by using a combined likelihood fit in the ten search regions of category A and B.
The theta tool~\cite{theta} is used to produce Bayesian limits on the signal cross section,
where the statistical and systematic uncertainties are treated as nuisance parameters.
Statistical and systematic uncertainties that are specific to category A or B, such as the uncertainties in the backgrounds from misidentified leptons, are assumed to be uncorrelated, whereas common sources of systematic uncertainties are treated as fully correlated.
The common uncertainties are the uncertainties in the jet energy scale and resolution, $\tau$ lepton and muon identification and isolation efficiencies, $\tau$ lepton energy scale and resolution, PDFs, and integrated luminosity.

\begin{figure}[htbp]
\centering
\includegraphics[height=0.45\textwidth]{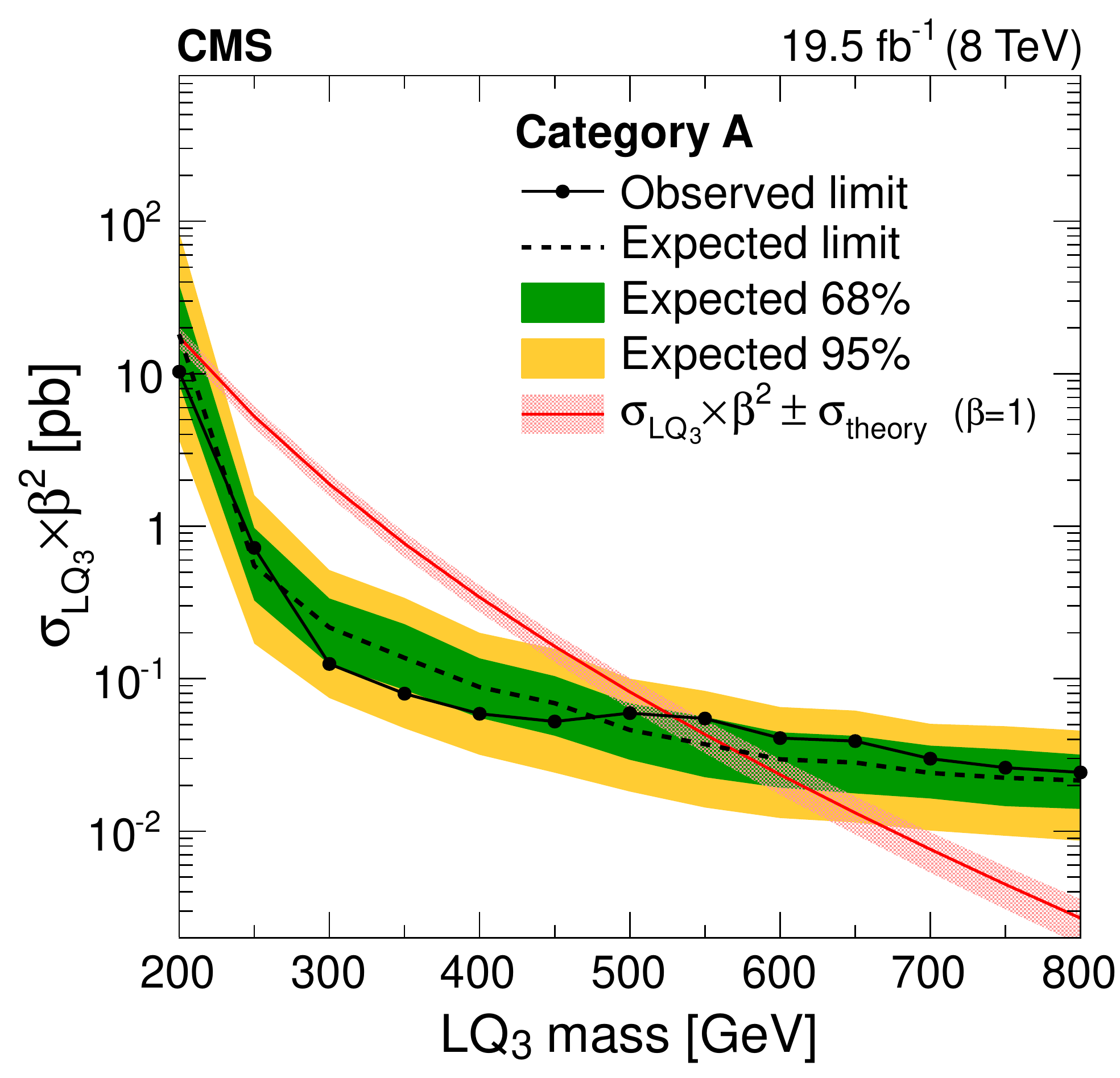}
\includegraphics[height=0.45\textwidth]{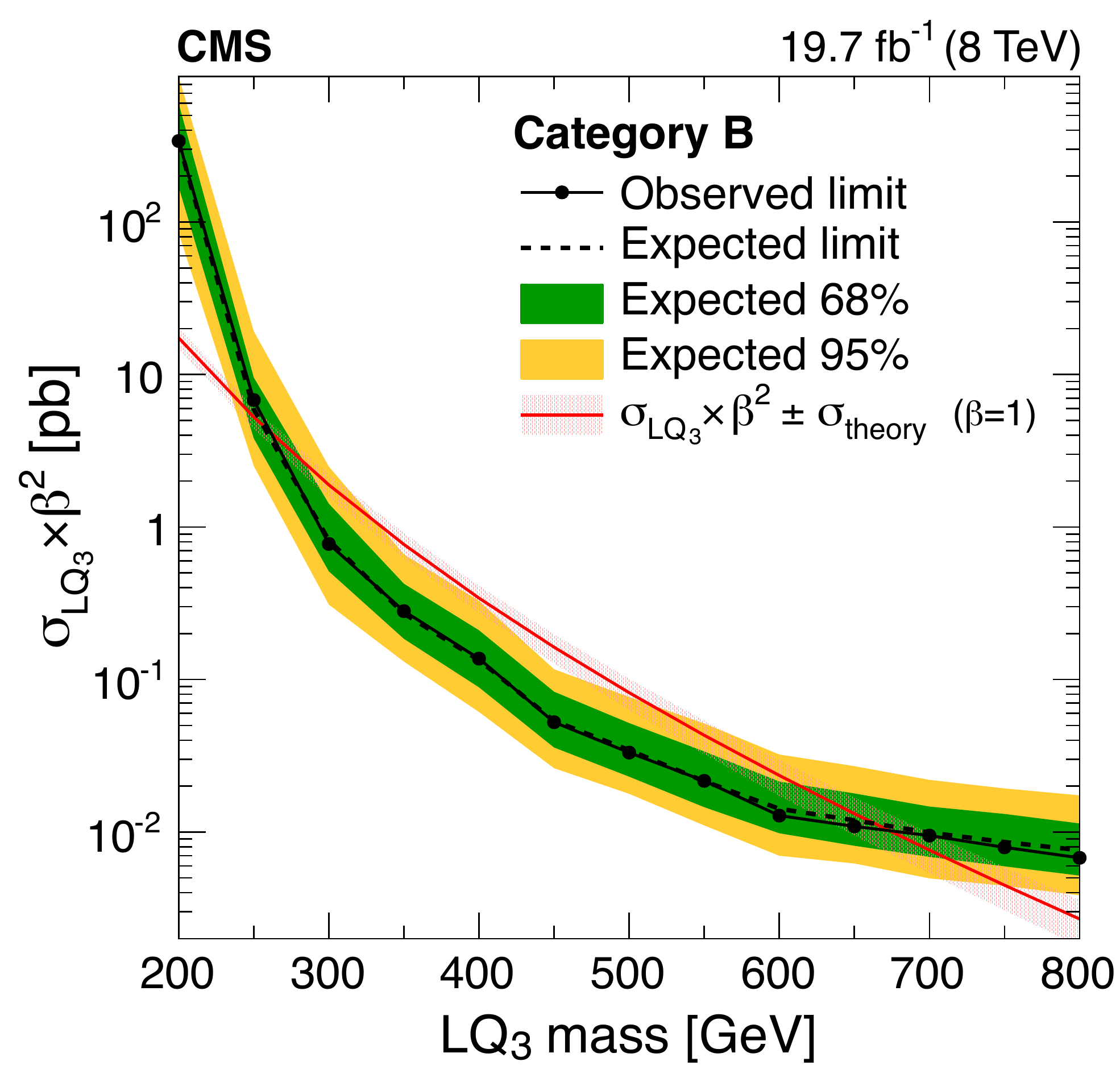}
\includegraphics[height=0.45\textwidth]{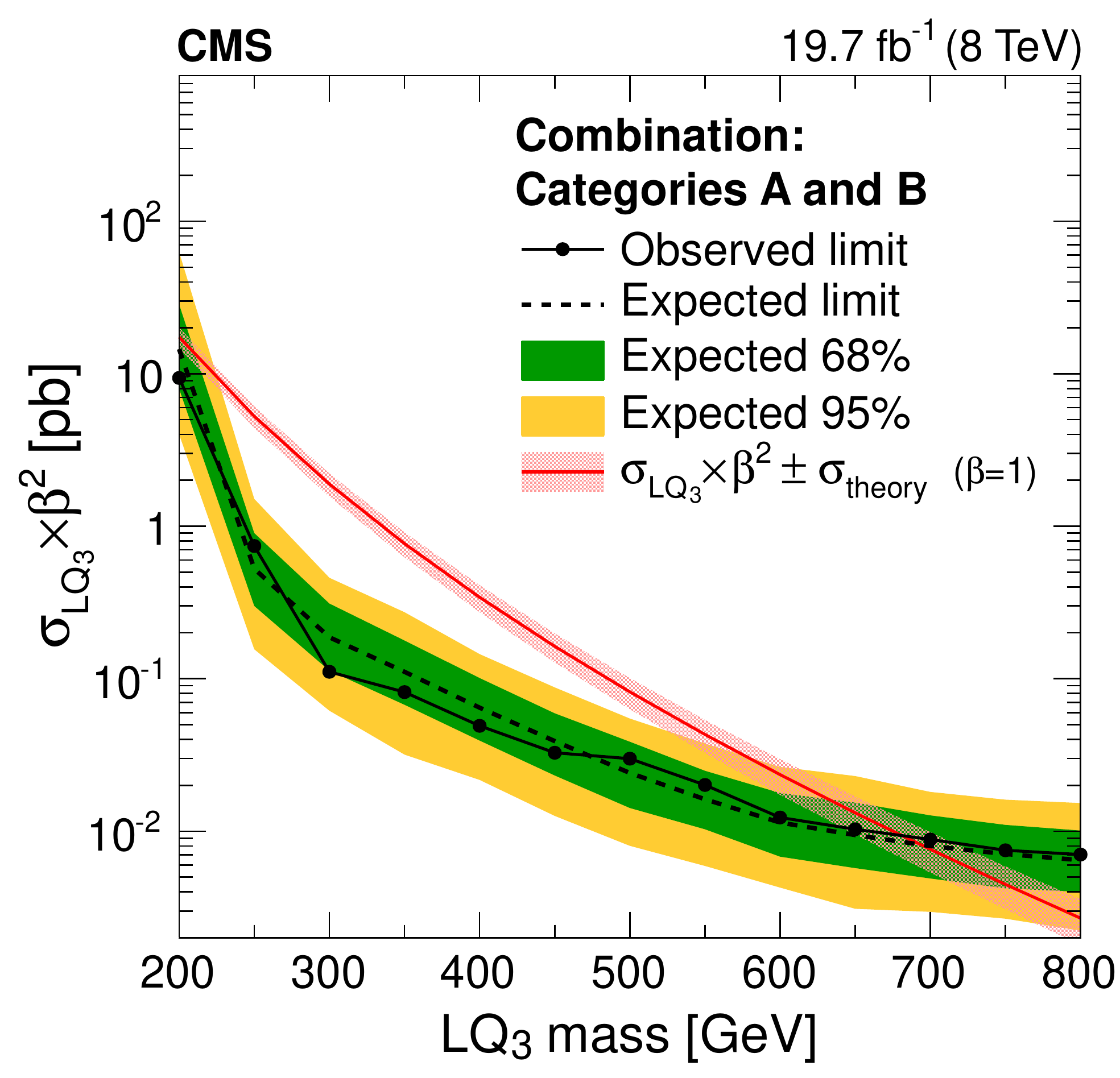}
\includegraphics[height=0.45\textwidth]{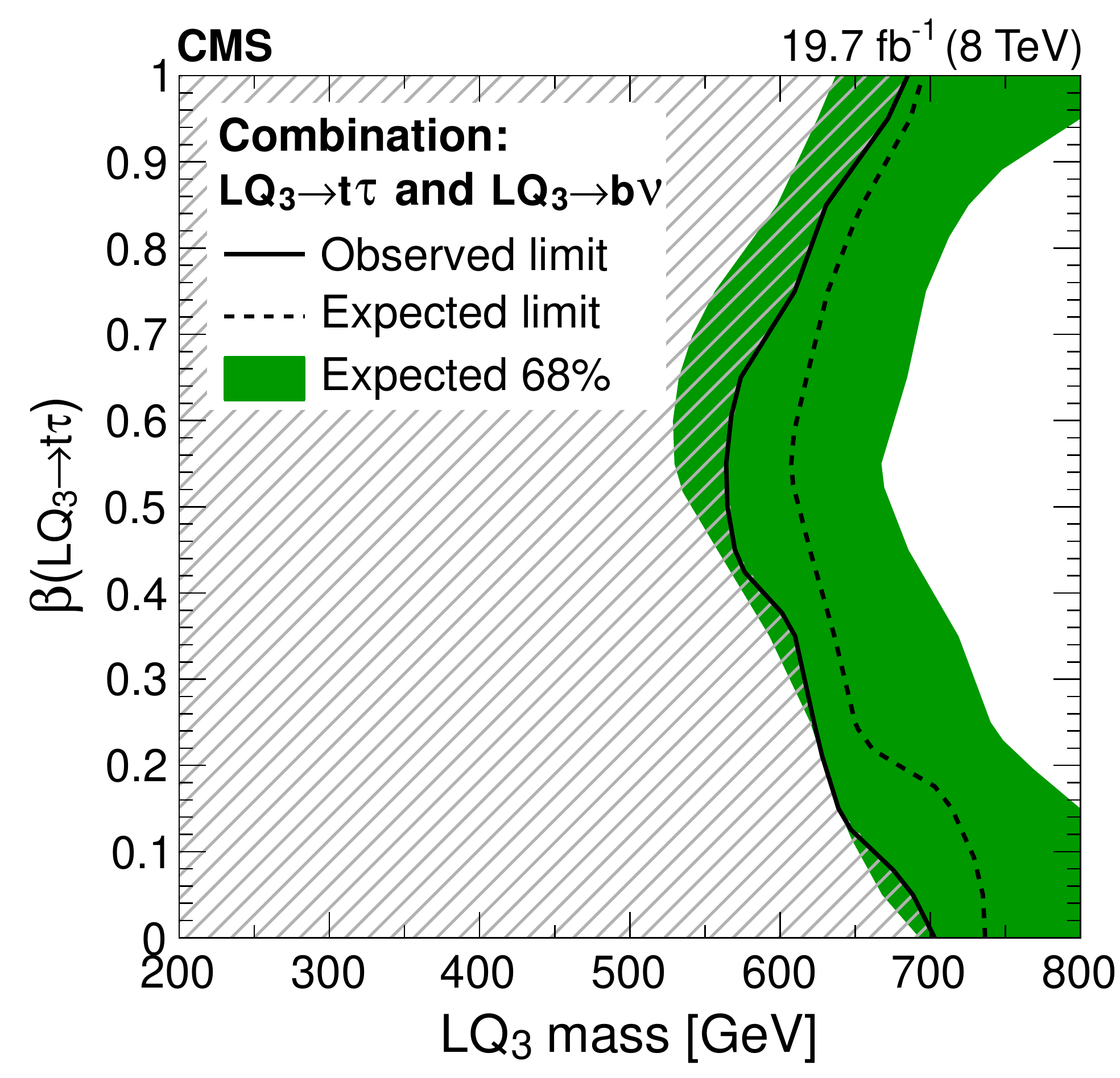}
{\caption{The expected and observed exclusion limits at 95\% CL on the $\LQt$ pair production cross section times $\beta^{2}$ in category A (upper left), category B (upper right) and the combination of the two categories (lower left).
The theoretical uncertainty in the LQ pair production cross section includes the PDF and renormalization/factorization scale uncertainties as prescribed in Ref.~\cite{Kramer}.
The expected and observed limits on the LQ branching fraction $\beta$ as a function of the LQ mass (lower right).
The total excluded region (shaded) is obtained by including the results in Ref.~\cite{Khachatryan:2015wza}, reinterpreted for the $\LQt\to \PQb\nu$ scenario.
}\label{DATAlimitPlots}}
\end{figure}

The observed and expected exclusion limits as a function of the LQ mass are shown in Fig.~\ref{DATAlimitPlots}.
Assuming a unit branching fraction of LQ decays to top quark and $\tau$ lepton pairs, pair production of third-generation LQs is excluded for masses up to 685\GeV with an expected limit of 695\GeV.
The exclusion limits worsen as the $\LQt$ mass approaches the mass of the top quark because the $\LQt$ decay products become softer.
At $M_{\LQt}=200$\GeV, more than 90\% of $\tau$ leptons originating from $\LQt$ decays have $\pt<60$\GeV,
which causes a decrease both in the signal selection efficiency and the discriminating performance of the $\tau$ lepton $\pt$ spectrum.
Therefore, no exclusion limits are quoted for masses below 200\GeV.

Branching fraction dependent exclusion limits are presented in Fig.~\ref{DATAlimitPlots} (lower right), where limits on the complementary $\LQt\to \PQb\nu$ ($\beta=0$) decay channel are also included.
The results for $\beta=0$ are obtained via reinterpretation of a search for pair produced bottom squarks~\cite{Khachatryan:2015wza} with subsequent decays into b quark and neutralino pairs, in the limit of vanishing neutralino masses.
In a statistical combination of this analysis with the search for bottom squarks, third-generation scalar LQs are also excluded for masses below 700\GeV for $\beta=0$ and for masses below 560\GeV over the full $\beta$ range.
If upper limits on $\beta$ are to be used to constrain the lepton-quark-LQ Yukawa couplings, $\lambda_{\PQb\nu}$ and $\lambda_{\PQt\tau}$, kinematic suppression factors that favor $\PQb\nu$ decay over the t$\tau$ have to be considered as well as the relative strengths of the two Yukawa couplings \cite{PhysRevLett.99.061801,PhysRevD.77.091105}.

Additionally, the results presented here for the third-generation scalar LQs are directly reinterpreted in the context of pair produced bottom squarks decaying into top quark and $\tau$ lepton pairs.
Thus, pair production of bottom squarks where the decay mode is dominated by the RPV coupling $\lambda^{\prime}_{333}$ is also excluded up to a bottom squark mass of 685\GeV.

\begin{table}[htbp]
\centering
\topcaption{Category A search results in the signal region for several $\LQt$ mass hypotheses.
The $\tau$ lepton $\pt$ and $\ST$ columns represent the optimized thresholds defined in Section \ref{SelectionCatA}.
The corresponding expected number of prompt-prompt and total background events, as well as the observed number of data events are listed as $N_{\text{Bkg}}^{\mathrm{PP}}$, total $N_{\text{Bkg}}^{\text{Exp}}$, and $N^{\text{Obs}}$.
The statistical and systematic uncertainties quoted in the expected number of background events are combinations of misidentified lepton and prompt-prompt components.
The $\epsilon_{\LQt}$ is the expected signal efficiency at a given $\LQt$ mass with respect to the total number of expected $\LQt$ signal events at $\sqrt{s}=8$\TeV with a $\mu\tauh$ pair of any charge combination.
No expected signal efficiency for $M_{\LQt}=200$\GeV is reported in the forward channel since the associated yield in the signal sample was measured to be zero.
}\label{SignalRegionResultsTable}
\resizebox{\linewidth}{!}{%
\begin{tabular}{ c c c | c c c r c c }
\hline
 $M_{\LQt}$      & $p^{\tau}_{\rm T}$ &  $\ST$  & $N_\text{Bkg}^\mathrm{PP}$ & Total $N_\text{Bkg}^\text{Exp}$ & $N^\text{Obs}$ & Z-score & $N_{\LQt}^\text{Exp}$ &  $\epsilon_{\LQt}$  \\
 (\GeVns) & (\GeVns) & (\GeVns) & $\pm\stat$ & $\pm\stat\pm\syst$ & & & $\pm\stat$ & (\%) \\
\hline
\rule[\baselineskip]{0pt}{2.5ex}& & & \multicolumn{6}{c}{ Central channel: $\widetilde{\abs{\eta}}<0.9$ }\\
200 & 35 & 410 & $8.5\pm1.0$ & $128\pm5\pm25$~~~         & 105 & $-1.0$~~ & ~~$53\pm21$ & 0.04         \\
250 & 35 & 410 & $8.5\pm1.0$ & $128\pm5\pm25$~~~         & 105 & $-1.0$~~ & $252\pm24$  & 0.58         \\
300 & 50 & 470 & $4.2\pm0.5$ & ~~$39.9\pm2.9\pm8.3$~~  &  27 & $-1.5$~~ & $153\pm11$  & 0.98         \\
350 & 50 & 490 & $4.0\pm0.5$ & ~~$34.6\pm2.7\pm7.1$~~  &  25 & $-1.2$~~ & $92.4\pm5.6$    & 1.45         \\
400 & 65 & 680 & $0.9\pm0.2$ & ~~~~$7.2\pm1.2\pm1.7$~~ &   4 & $-1.0$~~ & $28.4\pm2.1$    & 1.00         \\
450 & 65 & 700 & $0.8\pm0.2$ & ~~~~$6.3\pm1.1\pm1.6$~~ &   4 & $-0.8$~~ & $17.3\pm1.1$    & 1.27         \\
500 & 65 & 770 & $0.5\pm0.2$ & ~~~~$3.2\pm0.8\pm0.8$~~ &   4 & $+0.5$~~ & ~~$9.8\pm0.6$   & 1.43        \\
550 & 65 & 800 & $0.4\pm0.1$ & ~~~~$2.7\pm0.8\pm0.6$~~ &   4 & $+0.7$~~ & ~~$6.1\pm0.3$   & 1.71         \\
600 & 65 & 850 & $0.2\pm0.1$ & ~~~~$1.8\pm0.6\pm0.4$~~ &   3 & $+0.9$~~ & ~~$3.6\pm0.2$   & 1.85         \\
650 & 65 & 850 & $0.2\pm0.1$ & ~~~~$1.8\pm0.6\pm0.4$~~ &   3 & $+0.9$~~ & ~~$2.2\pm0.1$   & 1.99         \\
700 & 85 & 850 & $0.1\pm0.1$ & ~~~~$1.1\pm0.5\pm0.3$~~ &   2 & $+0.8$~~ & ~~$1.3\pm0.1$   & 2.02         \\
750 & 85 & 850 & $0.1\pm0.1$ & ~~~~$1.1\pm0.5\pm0.3$~~ &   2 & $+0.8$~~ & ~~$0.8\pm0.1$   & 2.20         \\
800 & 85 & 850 & $0.1\pm0.1$ & ~~~~$1.1\pm0.5\pm0.3$~~ &   2 & $+0.8$~~ & ~~$0.5\pm0.1$   & 2.80        \\
\hline
\rule[\baselineskip]{0pt}{2.5ex}& & & \multicolumn{6}{c}{ Forward channel: $\widetilde{\abs{\eta}}\ge0.9$ }\\
200 & 35 & 410 & $4.2\pm0.5$ & $72\pm4\pm15$~          & 87 & $+1.1$~~ &  ~~---        & ---          \\
250 & 35 & 410 & $4.2\pm0.5$ & $72\pm4\pm15$~          & 87 & $+1.1$~~ & ~~$50\pm11$    & 0.11          \\
300 & 50 & 470 & $1.8\pm0.3$ & ~~$20.3\pm2.2\pm3.9$~~  & 23 & $+0.5$~~ & $33.4\pm5.2$       & 0.21          \\
350 & 50 & 490 & $1.7\pm0.3$ & ~~$18.2\pm2.0\pm3.5$~~  & 19 & $+0.2$~~ & $18.5\pm2.5$       & 0.29          \\
400 & 65 & 680 & $0.7\pm0.2$ & ~~~~$2.7\pm0.7\pm0.6$~~ &  1 & $-0.9$~~ & ~~$6.1\pm1.0$      & 0.21          \\
450 & 65 & 700 & $0.7\pm0.2$ & ~~~~$2.3\pm0.6\pm0.4$~~ &  1 & $-0.7$~~ & ~~$3.8\pm0.5$      & 0.28          \\
500 & 65 & 770 & $0.5\pm0.1$ & ~~~~$1.2\pm0.4\pm0.2$~~ &  1 &    0.0~~ & ~~$1.6\pm0.2$      & 0.24    \\
550 & 65 & 800 & $0.4\pm0.1$ & ~~~~$0.9\pm0.4\pm0.2$~~ &  1 & $+0.3$~~ & ~~$1.2\pm0.2$      & 0.32          \\
600 & 65 & 850 & $0.3\pm0.1$ & ~~~~$0.6\pm0.3\pm0.1$~~ &  1 & $+0.6$~~ & ~~$0.6\pm0.1$      & 0.29          \\
650 & 65 & 850 & $0.3\pm0.1$ & ~~~~$0.6\pm0.3\pm0.1$~~ &  1 & $+0.6$~~ & ~~$0.3\pm0.1$      & 0.26          \\
700 & 85 & 850 & $0.1\pm0.1$ & ~~~~$0.4\pm0.2\pm0.1$~~ &  0 & $-0.4$~~ & ~~$0.2\pm0.1$      & 0.28          \\
750 & 85 & 850 & $0.1\pm0.1$ & ~~~~$0.4\pm0.2\pm0.1$~~ &  0 & $-0.4$~~ & ~~$0.1\pm0.1$      & 0.35          \\
800 & 85 & 850 & $0.1\pm0.1$ & ~~~~$0.4\pm0.2\pm0.1$~~ &  0 & $-0.4$~~ & ~~$0.1\pm0.1$      & 0.36          \\
\hline
\end{tabular}}
\end{table}

\begin{table}[htbp]
\centering
\topcaption{Category B search results for the four $\pt^\tau$ search regions of the $\mu\tauh$ channel.
All expected values for background and signal processes ($\LQt$ masses indicated in parentheses) are reported with the corresponding statistical and systematic uncertainties.
The expected signal efficiency $\epsilon_{\LQt}$ at a given $\LQt$ mass is determined with respect to the total number of expected $\LQt$ signal events at $\sqrt{s}=8$\TeV with a $\mu\tauh$ pair of any charge combination, and
$\epsilon_{\LQt}$ is reported separately for opposite-sign (OS) and same-sign (SS) $\mu\tauh$ events. }\label{SignalRegionResultsTableOS}
\resizebox{\linewidth}{!}{%
\renewcommand{\arraystretch}{1.3}
\begin{tabular}{ c | l l l l c c}
\hline
\multirow{2}{*}{ Process} & \multirow{2}{*}{~~$\pt^\tau<60$\GeV} & \multirow{2}{*}{$60 < \pt^\tau < 120$\GeV} & \multirow{2}{*}{$120 < \pt^\tau < 200$\GeV} & \multirow{2}{*}{$\pt^\tau > 200$\GeV} & \multicolumn{2}{c}{$\epsilon_{\LQt}$ (\%)} \\
 &  &  &  &  & OS &  SS \\
\hline
$\LQt$ (200\GeV) & ~~~~~$21\pm12^{+7}_{-2}$      & ~~~~$0.0\pm 0.1\pm0.0$        & ~~~~~$0.0\pm0.1\pm0.1$         & $0.0\pm0.1\pm0.1$         & 0.01 & 0   \\
$\LQt$ (250\GeV) & ~~$31.0\pm8.2^{+6.6}_{-3.4}$  & ~~$13.1\pm5.5^{+1.1}_{-2.9}$  & ~~~~~$0.0\pm0.1\pm0.1$         & $0.0\pm0.1\pm0.1$         & 0.09 & 0.02\\
$\LQt$ (300\GeV) & ~~$33.1\pm5.3^{+2.8}_{-3.8}$  & ~~$24.6\pm4.6^{+2.8}_{-2.1}$  & ~~~~~$7.6\pm2.6^{+1.1}_{-1.7}$ & $3.9\pm1.8^{+0.9}_{-0.3}$ & 0.35 & 0.08\\
$\LQt$ (350\GeV) & ~~$18.1\pm2.6^{+1.8}_{-1.4}$  & ~~$13.3\pm2.2^{+1.0}_{-1.1}$  & ~~~~~$7.2\pm1.6^{+0.8}_{-0.7}$ & $2.9\pm0.9^{+0.5}_{-1.4}$ & 0.57 & 0.08\\
$\LQt$ (400\GeV) & ~~$13.9\pm1.4^{+1.1}_{-2.6}$  & ~~$13.4\pm1.4^{+1.0}_{-1.1}$  & ~~~~~$7.8\pm1.1^{+0.8}_{-0.6}$ & $4.1\pm0.8^{+0.6}_{-0.8}$ & 1.30 & 0.12\\
$\LQt$ (450\GeV) & ~~$10.1\pm0.9^{+0.8}_{-1.9}$  & ~~~~$8.6\pm0.8^{+0.8}_{-0.8}$ & ~~~~~$7.1\pm0.7^{+0.5}_{-0.6}$ & $5.8\pm0.6^{+0.7}_{-0.6}$ & 2.05 & 0.27\\
$\LQt$ (500\GeV) & ~~~~$5.2\pm0.4^{+0.5}_{-0.9}$ & ~~~~$6.0\pm0.5\pm0.5$         & ~~~~~$5.3\pm0.4^{+0.4}_{-0.5}$ & $4.4\pm0.4^{+0.7}_{-0.5}$ & 2.75 & 0.27\\
$\LQt$ (550\GeV) & ~~~~$3.2\pm0.3^{+0.3}_{-0.6}$ & ~~~~$4.4\pm0.3^{+0.4}_{-0.3}$ & ~~~~~$4.3\pm0.3^{+0.5}_{-0.4}$ & $4.0\pm0.3\pm0.4$         & 4.04 & 0.36\\
$\LQt$ (600\GeV) & ~~~~$2.0\pm0.1^{+0.2}_{-0.5}$ & ~~~~$2.7\pm0.2\pm0.2$         & ~~~~~$2.7\pm0.2\pm0.2$         & $3.5\pm0.2\pm0.4$         & 5.11 & 0.43\\
$\LQt$ (650\GeV) & ~~~~$1.3\pm0.1^{+0.1}_{-0.3}$ & ~~~~$1.8\pm0.1^{+0.1}_{-0.2}$ & ~~~~~$2.0\pm0.1\pm 0.2$        & $2.5\pm0.1^{+0.3}_{-0.2}$ & 6.07 & 0.67\\
$\LQt$ (700\GeV) & ~~~~$0.7\pm0.1\pm0.1$         & ~~~~$1.1\pm0.1\pm0.1$         & ~~~~~$1.1\pm0.1\pm 0.1$        & $1.6\pm0.1^{+0.2}_{-0.1}$ & 6.66 & 0.57\\
$\LQt$ (750\GeV) & ~~~~$0.4\pm0.1\pm0.1$         & ~~~~$0.5\pm0.1\pm0.1$         & ~~~~~$0.7\pm0.1\pm0.1$         & $1.1\pm0.1\pm0.1$         & 6.71 & 0.59\\
$\LQt$ (800\GeV) & ~~~~$0.2\pm0.1\pm0.1$         & ~~~~$0.4\pm0.1\pm0.1$         & ~~~~~$0.5\pm0.1\pm0.1$         & $0.8\pm0.1\pm0.1$         & 7.77 & 0.61\\
\hline
$\ttbar$+jets & ~~$29.9\pm2.9^{+7.3}_{-7.2}$  & ~~~~$8.8\pm1.3^{+3.2}_{-3.4}$  & ~~~~~$1.7\pm0.6^{+0.6}_{-0.6}$ & $0.4\pm0.3^{+0.9}_{-0.4}$ \\
W+jets                & ~~~~$7.4\pm1.7^{+5.1}_{-5.1}$ & ~~~~$0.6\pm0.5\pm0.6$          & ~~~~~$0.0\pm0.1\pm0.1$         & $0.4\pm0.4\pm0.4$         \\
DY+jets               & ~~~~$4.8\pm0.7\pm2.5$         & ~~~~$1.8\pm0.4^{+1.1}_{-0.9}$  & ~~~~~$0.5\pm0.2\pm0.3$         & $0.4\pm0.2\pm0.2$         \\
Other backgrounds      & ~~~~$3.1\pm0.9^{+1.8}_{-1.9}$ & ~~~~$0.2\pm0.1^{+0.8}_{-0.3}$  & ~~~~~$0.2\pm0.1\pm0.4$         & $0.1\pm0.1^{+0.1}_{-0.2}$ \\
\hline
Total $N_\text{Bkg}^\text{Exp}$ & ~~$45.2\pm3.5^{+9.4}_{-9.3}$  & ~~$11.5\pm1.4^{+3.4}_{-3.6}$  & ~~~~~$2.5\pm0.6\pm0.8$  &  $1.2\pm0.5^{+1.0}_{-0.6}$ \\
$N^\text{Obs}$ & \multicolumn{1}{c}{44} & \multicolumn{1}{c}{15} & \multicolumn{1}{c}{1} & \multicolumn{1}{c}{0}\\
Z-score & \multicolumn{1}{c}{$-0.1~~$}   &\multicolumn{1}{c}{$+0.7~~$}  & \multicolumn{1}{c}{$+0.8~~$} & \multicolumn{1}{c}{$-1.0~~$} & \\
\hline\end{tabular}}
\end{table}

\begin{table}[htbp]
\centering
\topcaption{Category B search results for the four $\pt^\tau$ search regions of the $\Pe\tauh$ channel.
All expected values for background and signal processes ($\LQt$ masses indicated in parentheses) are reported with the corresponding statistical and systematic uncertainties.
The expected signal efficiency $\epsilon_{\LQt}$ at a given $\LQt$ mass is determined with respect to the total number of expected $\LQt$ signal events at $\sqrt{s}=8$\TeV with an $\Pe\tauh$ pair of any charge combination.}
\label{SignalRegionResultsTableOSele}
\resizebox{\linewidth}{!}{%
\renewcommand{\arraystretch}{1.3}
\begin{tabular}{ c | l l l l c}
\hline
 Process & ~~$\pt^\tau < 60$\GeV & $60 < \pt^\tau < 120$\GeV & $120 < \pt^\tau < 200$\GeV & $\pt^\tau > 200$\GeV & $\epsilon_{\LQt}$ \\
\hline
$\LQt$ (200\GeV) & ~~~~~$32\pm19^{+6}_{-4}$      & ~~~~~$0.1\pm0.1\pm0.1$         & ~~~~~$0.0\pm0.1\pm0.1$         & $0.0\pm0.1\pm0.1$ & 0.02\\
$\LQt$ (250\GeV) & ~~$33.3\pm8.7^{+8.2}_{-5.8}$  & ~~~$11.9\pm5.3^{+2.6}_{-2.4}$  & ~~~~~$0.0\pm0.1\pm0.1$         & $0.0\pm0.1\pm0.1$ & 0.16\\
$\LQt$ (300\GeV) & ~~$31.9\pm5.2^{+3.2}_{-9.1}$  & ~~~$27.7\pm4.6^{+3.9}_{-4.4}$  & ~~~~~$4.2\pm1.9^{+1.1}_{-0.3}$ & $2.7\pm1.6^{+0.2}_{-0.3}$ & 0.70\\
$\LQt$ (350\GeV) & ~~$19.6\pm2.6^{+2.7}_{-3.6}$  & ~~~$19.6\pm2.5^{+2.0}_{-2.1}$  & ~~~~~$8.6\pm1.7^{+1.4}_{-0.8}$ & $4.7\pm1.3\pm0.5$ & 1.35\\
$\LQt$ (400\GeV) & ~~$12.7\pm1.4^{+1.7}_{-2.6}$  & ~~~$14.6\pm1.5^{+2.0}_{-1.4}$  & ~~~~~$8.1\pm1.1^{+1.2}_{-1.3}$ & $4.8\pm0.9^{+0.6}_{-0.4}$ & 2.22\\
$\LQt$ (450\GeV) & ~~~~$7.8\pm0.7^{+1.3}_{-1.4}$ & ~~~$10.1\pm0.9^{+0.9}_{-1.1}$  & ~~~~~$7.2\pm0.7^{+1.0}_{-0.7}$ & $5.5\pm0.6^{+0.6}_{-0.8}$ & 3.65\\
$\LQt$ (500\GeV) & ~~~~$4.8\pm0.4^{+0.5}_{-1.2}$ & ~~~~~$7.3\pm0.5^{+0.8}_{-0.9}$ & ~~~~~$5.5\pm0.4\pm0.6$         & $5.2\pm0.4^{+0.7}_{-0.6}$&5.34 \\
$\LQt$ (550\GeV) & ~~~~$3.3\pm0.2^{+0.4}_{-1.0}$ & ~~~~~$4.3\pm0.3\pm0.4$         & ~~~~~$4.4\pm0.3\pm0.4$         & $4.3\pm0.3\pm0.5$ & 7.28\\
$\LQt$ (600\GeV) & ~~~~$1.9\pm0.1^{+0.2}_{-0.6}$ & ~~~~~$2.9\pm0.2\pm0.3$         & ~~~~~$3.2\pm0.2\pm0.3$         & $3.6\pm0.2\pm0.4$ & 9.61\\
$\LQt$ (650\GeV) & ~~~~$1.2\pm0.1^{+0.1}_{-0.4}$ & ~~~~~$1.8\pm0.1\pm0.2$         & ~~~~~$2.0\pm0.1\pm0.2$         & $2.4\pm0.1^{+0.3}_{-0.3}$& 10.89 \\
$\LQt$ (700\GeV) & ~~~~$0.7\pm0.1^{+0.1}_{-0.2}$ & ~~~~~$1.1\pm0.1\pm0.1$         & ~~~~~$1.5\pm0.1\pm0.1$         & $1.9\pm0.1\pm0.2$ & 13.11\\
$\LQt$ (750\GeV) & ~~~~$0.4\pm0.1\pm0.1$         & ~~~~~$0.7\pm0.1\pm0.1$         & ~~~~~$0.7\pm0.1\pm0.1$         & $1.4\pm0.1^{+0.1}_{-0.2}$&13.84 \\
$\LQt$ (800\GeV) & ~~~~$0.2\pm0.1\pm0.1$         & ~~~~~$0.4\pm0.1\pm0.1$         & ~~~~~$0.5\pm0.1\pm0.1$         & $0.9\pm0.1\pm0.1$ &14.82\\
\hline
$\ttbar$+jets & ~~$27.7\pm2.4\pm7.7$          & ~~~~~$7.5\pm1.2^{+2.1}_{-2.8}$ & ~~~~~$0.9\pm0.4\pm0.3$ & $0.1\pm0.1^{+0.6}_{-0.1}$ & \\
W+jets                & ~~~~$8.5\pm1.8^{+5.3}_{-5.4}$ & ~~~~~$1.1\pm0.6^{+0.6}_{-0.7}$ & ~~~~~$0.0\pm0.1\pm0.1$ & $0.0\pm0.1\pm0.1$ & \\
DY+jets               & ~~~~$4.4\pm0.7^{+2.3}_{-2.4}$ & ~~~~~$1.4\pm0.4^{+0.8}_{-0.7}$ & ~~~~~$0.6\pm0.2\pm0.3$ & $0.2\pm0.1\pm0.1$ & \\
Other backgrounds     & ~~~~$3.5\pm1.0^{+2.5}_{-3.0}$ & ~~~~~$1.1\pm0.5^{+0.7}_{-0.8}$ & ~~~~~$0.2\pm0.1\pm0.2$ & $0.4\pm0.3\pm0.2$ & \\
\hline
Total $N_\text{Bkg}^\text{Exp}$ & ~~$44.1\pm3.2^{+10.0}_{-10.1}$ & ~~~$11.0\pm1.5^{+2.5}_{-3.1}$ & ~~~~~$1.6\pm0.5\pm0.5$ & $0.7\pm0.4^{+0.7}_{-0.3}$ \\
& & & & &\\[-2ex]
$N^\text{Obs}$ & \multicolumn{1}{c}{53} & \multicolumn{1}{c}{5} & \multicolumn{1}{c}{4} &\multicolumn{1}{c}{1}\\
Z-score & \multicolumn{1}{c}{$+0.8~~$} & \multicolumn{1}{c}{$-1.2~~$} & \multicolumn{1}{c}{$+1.1~~$} & \multicolumn{1}{c}{$+0.1~~$} & \\
\hline
\end{tabular}}
\end{table}

\clearpage
\section{Summary \label{Conclusions}}

A search for pair produced, charge $-1/3$, third-generation scalar leptoquarks decaying to top quark and $\tau$ lepton pairs has been conducted in the $\ell\tauh$ channel with two or more jets, using a proton-proton collisions data sample collected with the CMS detector at $\sqrt{s}=8$\TeV corresponding to an integrated luminosity of 19.7\fbinv.
No statistically significant excess is observed over the SM background expectations.
Assuming that all leptoquarks decay to a top quark and a $\tau$ lepton, the pair production of charge $-1/3$, third-generation scalar leptoquarks is excluded at 95\% CL for masses up to 685\GeV (695\GeV expected).
This constitutes the first direct result for leptoquarks decaying in this channel, and the mass limit is also directly applicable to pair produced bottom squarks decaying via the RPV coupling $\lambda_{333}^{\prime}$.

\begin{acknowledgments}
We congratulate our colleagues in the CERN accelerator departments for the excellent performance of the LHC and thank the technical and administrative staffs at CERN and at other CMS institutes for their contributions to the success of the CMS effort. In addition, we gratefully acknowledge the computing centers and personnel of the Worldwide LHC Computing Grid for delivering so effectively the computing infrastructure essential to our analyses. Finally, we acknowledge the enduring support for the construction and operation of the LHC and the CMS detector provided by the following funding agencies: BMWFW and FWF (Austria); FNRS and FWO (Belgium); CNPq, CAPES, FAPERJ, and FAPESP (Brazil); MES (Bulgaria); CERN; CAS, MoST, and NSFC (China); COLCIENCIAS (Colombia); MSES and CSF (Croatia); RPF (Cyprus); MoER, ERC IUT and ERDF (Estonia); Academy of Finland, MEC, and HIP (Finland); CEA and CNRS/IN2P3 (France); BMBF, DFG, and HGF (Germany); GSRT (Greece); OTKA and NIH (Hungary); DAE and DST (India); IPM (Iran); SFI (Ireland); INFN (Italy); MSIP and NRF (Republic of Korea); LAS (Lithuania); MOE and UM (Malaysia); CINVESTAV, CONACYT, SEP, and UASLP-FAI (Mexico); MBIE (New Zealand); PAEC (Pakistan); MSHE and NSC (Poland); FCT (Portugal); JINR (Dubna); MON, RosAtom, RAS and RFBR (Russia); MESTD (Serbia); SEIDI and CPAN (Spain); Swiss Funding Agencies (Switzerland); MST (Taipei); ThEPCenter, IPST, STAR and NSTDA (Thailand); TUBITAK and TAEK (Turkey); NASU and SFFR (Ukraine); STFC (United Kingdom); DOE and NSF (USA).

Individuals have received support from the Marie-Curie program and the European Research Council and EPLANET (European Union); the Leventis Foundation; the A. P. Sloan Foundation; the Alexander von Humboldt Foundation; the Belgian Federal Science Policy Office; the Fonds pour la Formation \`a la Recherche dans l'Industrie et dans l'Agriculture (FRIA-Belgium); the Agentschap voor Innovatie door Wetenschap en Technologie (IWT-Belgium); the Ministry of Education, Youth and Sports (MEYS) of the Czech Republic; the Council of Science and Industrial Research, India; the HOMING PLUS program of the Foundation for Polish Science, cofinanced from European Union, Regional Development Fund; the Compagnia di San Paolo (Torino); the Consorzio per la Fisica (Trieste); MIUR project 20108T4XTM (Italy); the Thalis and Aristeia programs cofinanced by EU-ESF and the Greek NSRF; and the National Priorities Research Program by Qatar National Research Fund.
\end{acknowledgments}

\bibliography{auto_generated}

\providecommand{\href}[2]{#2}\begingroup\raggedright\begin{thebibliography}{10}%
\makeatletter
\providecommand{\hrefCMSnoop }[0]{\@secondoftwo}%
\makeatother
\providecommand{\doi}{\texttt{doi:}\begingroup \urlstyle{tt}\Url}

\bibitem{PatiSalam}
\hrefCMSnoop {}{``Lepton number as the fourth ``color'',}
  \href{http://dx.doi.org/10.1103/PhysRevD.10.275}{\doi{10.1103/PhysRevD.10.275}}.
  [Erratum: \DOI{10.1103/PhysRevD.11.703.2}].

\bibitem{GeorgiGlashow}
\hrefCMSnoop {}{H.~Georgi and S.~L. Glashow, ``{Unity of All Elementary
  Particle Forces}'',} \textit{ Phys. Rev. Lett.} \textbf{ 32} (1974) 438,
  \href{http://dx.doi.org/10.1103/PhysRevLett.32.438}{\doi{10.1103/PhysRevLett.32.438}}.

\bibitem{TOPSU(5)Model}
\hrefCMSnoop {}{S.~Chakdar, T.~Li, S.~Nandi, and S.~K. Rai, ``{Unity of
  elementary particles and forces for the third family}'',} \textit{ Phys.
  Lett. B} \textbf{ 718} (2012) 121,
  \href{http://dx.doi.org/10.1016/j.physletb.2012.10.021}{\doi{10.1016/j.physletb.2012.10.021}},
  \href{http://www.arXiv.org/abs/1206.0409}{\texttt{arXiv:1206.0409}}.

\bibitem{Technicolor}
\hrefCMSnoop {}{E.~Farhi and L.~Susskind, ``{Technicolor}'',} \textit{ Phys.
  Rept.} \textbf{ 74} (1981) 277,
  \href{http://dx.doi.org/10.1016/0370-1573(81)90173-3}{\doi{10.1016/0370-1573(81)90173-3}}.

\bibitem{LightLeptoquarks}
\hrefCMSnoop {}{B.~Schrempp and F.~Schrempp, ``Light leptoquarks'',} \textit{
  Phys. Lett. B} \textbf{ 153} (1985) 101,
  \href{http://dx.doi.org/10.1016/0370-2693(85)91450-9}{\doi{10.1016/0370-2693(85)91450-9}}.

\bibitem{Gripaios:2009dq}
\hrefCMSnoop {}{B.~Gripaios, ``Composite leptoquarks at the {LHC}'',} \textit{
  JHEP} \textbf{ 02} (2010) 045,
  \href{http://dx.doi.org/10.1007/JHEP02(2010)045}{\doi{10.1007/JHEP02(2010)045}},
  \href{http://www.arXiv.org/abs/0910.1789}{\texttt{arXiv:0910.1789}}.

\bibitem{Buchmuller1987442}
\hrefCMSnoop {}{W.~Buchm{\"u}ller, R.~R{\"u}ckl, and D.~Wyler, ``Leptoquarks in
  lepton-quark collisions'',} \textit{ Phys. Lett. B} \textbf{ 191} (1987) 442,
  \href{http://dx.doi.org/10.1016/0370-2693(87)90637-X}{\doi{10.1016/0370-2693(87)90637-X}}.
  [Erratum: \DOI{10.1016/S0370-2693(99)00014-3}].

\bibitem{LQsearchesAtHeraTevatron}
\hrefCMSnoop {}{D.~E. Acosta and S.~K. Blessing, ``{Leptoquark Searches at HERA
  and the Tevatron}'',} \textit{ Ann. Rev. Nucl. Part. Sci.} \textbf{ 49}
  (1999) 389,
  \href{http://dx.doi.org/10.1146/annurev.nucl.49.1.389}{\doi{10.1146/annurev.nucl.49.1.389}}.

\bibitem{Aad:2011ch}
\hrefCMSnoop {}{{ATLAS Collaboration}, ``{Search for first generation scalar
  leptoquarks in pp collisions at $\sqrt{s}=7$ TeV with the ATLAS detector}'',}
  \textit{ Phys. Lett. B} \textbf{ 709} (2012) 158,
  \href{http://dx.doi.org/10.1016/j.physletb.2012.02.004}{\doi{10.1016/j.physletb.2012.02.004}},
  \href{http://www.arXiv.org/abs/1112.4828}{\texttt{arXiv:1112.4828}}.
[Erratum: \DOI{10.1016/j.physletb.2012.03.023}].

\bibitem{ATLAS:2012aq}
\hrefCMSnoop {}{{ATLAS Collaboration}, ``{Search for second generation scalar
  leptoquarks in pp collisions at $\sqrt{s}=7$ TeV with the ATLAS detector}'',}
  \textit{ Eur. Phys. J. C} \textbf{ 72} (2012) 2151,
  \href{http://dx.doi.org/10.1140/epjc/s10052-012-2151-6}{\doi{10.1140/epjc/s10052-012-2151-6}},
\href{http://www.arXiv.org/abs/1203.3172}{\texttt{arXiv:1203.3172}}.

\bibitem{Chatrchyan:2012vza}
\hrefCMSnoop {}{{CMS Collaboration}, ``{Search for pair production of first-
  and second-generation scalar leptoquarks in pp collisions at $\sqrt{s}= 7$
  TeV}'',} \textit{ Phys. Rev. D} \textbf{ 86} (2012) 052013,
  \href{http://dx.doi.org/10.1103/PhysRevD.86.052013}{\doi{10.1103/PhysRevD.86.052013}},
\href{http://www.arXiv.org/abs/1207.5406}{\texttt{arXiv:1207.5406}}.

\bibitem{PhysRevLett.99.061801}
\hrefCMSnoop {}{{D0} Collaboration, ``Search for Third-Generation Scalar
  Leptoquarks in $\rm{p}\overline{\rm{p}}$ Collisions at
  $\sqrt{s}=1.96$\,TeV'',} \textit{ Phys. Rev. Lett.} \textbf{ 99} (2007)
  061801,
  \href{http://dx.doi.org/10.1103/PhysRevLett.99.061801}{\doi{10.1103/PhysRevLett.99.061801}},
\href{http://www.arXiv.org/abs/0705.0812}{\texttt{arXiv:0705.0812}}.

\bibitem{PhysRevD.77.091105}
\hrefCMSnoop {}{{CDF} Collaboration, ``Search for third generation vector
  leptoquarks in $\rm{p}\overline{\rm{p}}$ collisions at
  $\sqrt{s}=1.96\text{\,}\text{\,}\mathrm{TeV}$'',} \textit{ Phys. Rev. D}
  \textbf{ 77} (2008) 091105,
  \href{http://dx.doi.org/10.1103/PhysRevD.77.091105}{\doi{10.1103/PhysRevD.77.091105}},
\href{http://www.arXiv.org/abs/0706.2832}{\texttt{arXiv:0706.2832}}.

\bibitem{ATLAS:2013oea}
\hrefCMSnoop {}{{ATLAS Collaboration}, ``{Search for third generation scalar
  leptoquarks in pp collisions at $\sqrt{s}$ = 7 TeV with the ATLAS
  detector}'',} \textit{ JHEP} \textbf{ 06} (2013) 033,
  \href{http://dx.doi.org/10.1007/JHEP06(2013)033}{\doi{10.1007/JHEP06(2013)033}},
\href{http://www.arXiv.org/abs/1303.0526}{\texttt{arXiv:1303.0526}}.

\bibitem{Chatrchyan:2012st}
\hrefCMSnoop {}{{CMS Collaboration}, ``{Search for third-generation leptoquarks
  and scalar bottom quarks in pp collisions at $\sqrt{s}=7$ TeV}'',} \textit{
  JHEP} \textbf{ 12} (2012) 055,
  \href{http://dx.doi.org/10.1007/JHEP12(2012)055}{\doi{10.1007/JHEP12(2012)055}},
  \href{http://www.arXiv.org/abs/1210.5627}{\texttt{arXiv:1210.5627}}.

\bibitem{Khachatryan:2014ura}
\hrefCMSnoop {}{{CMS Collaboration}, ``{Search for pair production of
  third-generation scalar leptoquarks and top squarks in proton-proton
  collisions at $\sqrt{s}$ = 8 TeV}'',} \textit{ Phys. Lett. B} \textbf{ 739}
  (2014) 229,
  \href{http://dx.doi.org/10.1016/j.physletb.2014.10.063}{\doi{10.1016/j.physletb.2014.10.063}},
\href{http://www.arXiv.org/abs/1408.0806}{\texttt{arXiv:1408.0806}}.

\bibitem{Khachatryan:2015wza}
\hrefCMSnoop {}{{CMS Collaboration}, ``{Searches for third generation squark
  production in fully hadronic final states in proton-proton collisions at
  $\sqrt{s}=8$ TeV}'',} (2015).
  \href{http://www.arXiv.org/abs/1503.08037}{\texttt{arXiv:1503.08037}}.
Submitted to \textit{JHEP}.

\bibitem{Barbier:2004ez}
R.~Barbier\hrefCMSnoop {}{ {et~al.}, ``{$R$}-Parity-violating supersymmetry'',}
  \textit{ Phys. Rept.} \textbf{ 420} (2005) 1,
  \href{http://dx.doi.org/10.1016/j.physrep.2005.08.006}{\doi{10.1016/j.physrep.2005.08.006}},
  \href{http://www.arXiv.org/abs/hep-ph/0406039}{\texttt{arXiv:hep-ph/0406039}}.

\bibitem{Kramer}
\hrefCMSnoop {}{M.~Kr{\"a}mer, T.~Plehn, M.~Spira, and P.~M. Zerwas, ``Pair
  production of scalar leptoquarks at the {CERN LHC}'',} \textit{ Phys. Rev. D}
  \textbf{ 71} (2005) 057503,
  \href{http://dx.doi.org/10.1103/PhysRevD.71.057503}{\doi{10.1103/PhysRevD.71.057503}},
\href{http://www.arXiv.org/abs/hep-ph/0411038}{\texttt{arXiv:hep-ph/0411038}}.

\bibitem{CMSdetector}
\hrefCMSnoop {}{{CMS Collaboration}, ``The {CMS} experiment at the {CERN}
  {LHC}'',} \textit{ JINST} \textbf{ 3} (2008) S08004,
\href{http://dx.doi.org/10.1088/1748-0221/3/08/S08004}{\doi{10.1088/1748-0221/3/08/S08004}}.

\bibitem{CMS:2009nxa}
\href {http://cdsweb.cern.ch/record/1194487}{{CMS Collaboration},
  ``{Particle-Flow Event Reconstruction in CMS and Performance for Jets, Taus,
  and $\ETmiss$}'',} CMS Physics Analysis Summary CMS-PAS-PFT-09-001, 2009.

\bibitem{CMS-PAS-PFT-10-001}
\href {http://cdsweb.cern.ch/record/1247373}{{CMS Collaboration},
  ``{Commissioning of the Particle-flow Event Reconstruction with the first LHC
  collisions recorded in the CMS detector}'',} CMS Physics Analysis Summary
  CMS-PAS-PFT-10-001, 2010.

\bibitem{Chatrchyan:2012xi}
\hrefCMSnoop {}{{CMS Collaboration}, ``{Performance of CMS muon reconstruction
  in pp collision events at $\sqrt{s}=7$ TeV}'',} \textit{ JINST} \textbf{ 7}
  (2012) P10002,
  \href{http://dx.doi.org/10.1088/1748-0221/7/10/P10002}{\doi{10.1088/1748-0221/7/10/P10002}},
\href{http://www.arXiv.org/abs/1206.4071}{\texttt{arXiv:1206.4071}}.

\bibitem{CMS-TAU-paper}
\hrefCMSnoop {}{{CMS Collaboration}, ``{Performance of tau-lepton
  reconstruction and identification in CMS}'',} \textit{ JINST} \textbf{ 7}
  (2012) P01001,
  \href{http://dx.doi.org/10.1088/1748-0221/7/01/P01001}{\doi{10.1088/1748-0221/7/01/P01001}},
  \href{http://www.arXiv.org/abs/1109.6034}{\texttt{arXiv:1109.6034}}.

\bibitem{Khachatryan:2015hwa}
\hrefCMSnoop {}{{CMS Collaboration}, ``{Performance of electron reconstruction
  and selection with the CMS detector in proton-proton collisions at
  $\sqrt{s}=8$ TeV}'',} (2015).
  \href{http://www.arXiv.org/abs/1502.02701}{\texttt{arXiv:1502.02701}}.
Submitted to \textit{JINST}.

\bibitem{antiKT}
\hrefCMSnoop {}{M.~Cacciari, G.~P. Salam, and G.~Soyez, ``The anti-$k_t$ jet
  clustering algorithm'',} \textit{ JHEP} \textbf{ 04} (2008) 063,
  \href{http://dx.doi.org/10.1088/1126-6708/2008/04/063}{\doi{10.1088/1126-6708/2008/04/063}}.

\bibitem{JEScalib}
\hrefCMSnoop {}{{CMS Collaboration}, ``Determination of jet energy calibration
  and transverse momentum resolution in {CMS}'',} \textit{ J. Instrum.}
  \textbf{ 6} (2011) P11002,
  \href{http://dx.doi.org/10.1088/1748-0221/6/11/P11002}{\doi{10.1088/1748-0221/6/11/P11002}}.

\bibitem{CMS-PAS-JME-12-002}
\href {http://cdsweb.cern.ch/record/1543527}{{CMS Collaboration},
  ``{Performance of Missing Transverse Momentum Reconstruction Algorithms in
  Proton-Proton Collisions at $\sqrt{s}=8$ TeV with the CMS Detector}'',} CMS
  Physics Analysis Summary CMS-PAS-JME-12-002, 2013.

\bibitem{Adam:2005zf}
\hrefCMSnoop {}{{CMS Collaboration}, ``{The CMS high level trigger}'',}
  \textit{ Eur. Phys. J. C} \textbf{ 46} (2006) 605,
  \href{http://dx.doi.org/10.1140/epjc/s2006-02495-8}{\doi{10.1140/epjc/s2006-02495-8}},
  \href{http://www.arXiv.org/abs/hep-ex/0512077}{\texttt{arXiv:hep-ex/0512077}}.

\bibitem{pythia}
\hrefCMSnoop {}{T.~Sj\, ``PYTHIA 6.4 physics and manual'',} \textit{ JHEP}
\href{http://dx.doi.org/10.1088/1126-6708/2006/05/026}{\doi{10.1088/1126-6708/2006/05/026}}.

\bibitem{powheg1}
\hrefCMSnoop {}{S.~Alioli, P.~Nason, C.~Oleari, and E.~Re, ``{A general
  framework for implementing NLO calculations in shower Monte Carlo programs:
  the POWHEG BOX}'',} \textit{ JHEP} \textbf{ 06} (2010) 043,
  \href{http://dx.doi.org/10.1007/JHEP06(2010)043}{\doi{10.1007/JHEP06(2010)043}},
  \href{http://www.arXiv.org/abs/1002.2581}{\texttt{arXiv:1002.2581}}.

\bibitem{Alioli:2009je}
\hrefCMSnoop {}{S.~Alioli, P.~Nason, C.~Oleari, and E.~Re, ``{NLO single-top
  production matched with shower in POWHEG: $s$- and $t$-channel
  contributions}'',} \textit{ JHEP} \textbf{ 09} (2009) 111,
  \href{http://dx.doi.org/10.1088/JHEP09(2009)111}{\doi{10.1088/JHEP09(2009)111}},
  \href{http://www.arXiv.org/abs/0907.4076}{\texttt{arXiv:0907.4076}}.
[Erratum: \DOI{10.1007/JHEP02(2010)011}].

\bibitem{Re:2010bp}
\hrefCMSnoop {}{E.~Re, ``{Single-top Wt-channel production matched with parton
  showers using the POWHEG method}'',} \textit{ Eur. Phys. J. C} \textbf{ 71}
  (2011) 1547,
  \href{http://dx.doi.org/10.1140/epjc/s10052-011-1547-z}{\doi{10.1140/epjc/s10052-011-1547-z}},
\href{http://www.arXiv.org/abs/1009.2450}{\texttt{arXiv:1009.2450}}.

\bibitem{Frixione:2007nw}
\hrefCMSnoop {}{S.~Frixione, P.~Nason, and G.~Ridolfi, ``{A Positive-weight
  next-to-leading-order Monte Carlo for heavy flavour hadroproduction}'',}
  \textit{ JHEP} \textbf{ 09} (2007) 126,
  \href{http://dx.doi.org/10.1088/1126-6708/2007/09/126}{\doi{10.1088/1126-6708/2007/09/126}},
\href{http://www.arXiv.org/abs/0707.3088}{\texttt{arXiv:0707.3088}}.

\bibitem{madgraph}
J.~Alwall\hrefCMSnoop {}{ {et~al.}, ``MadGraph 5: going beyond'',} \textit{
  JHEP} \textbf{ 06} (2011) 128,
  \href{http://dx.doi.org/10.1007/JHEP06(2011)128}{\doi{10.1007/JHEP06(2011)128}},
\href{http://www.arXiv.org/abs/1106.0522}{\texttt{arXiv:1106.0522}}.

\bibitem{MLM}
\hrefCMSnoop {}{J.~Alwall {et~al.}, ``{Comparative study of various algorithms
  for the merging of parton showers and matrix elements in hadronic
  collisions}'',} \textit{ Eur. Phys. J. C} \textbf{ 53} (2008) 473,
  \href{http://dx.doi.org/10.1140/epjc/s10052-007-0490-5}{\doi{10.1140/epjc/s10052-007-0490-5}},
\href{http://www.arXiv.org/abs/0706.2569}{\texttt{arXiv:0706.2569}}.

\bibitem{tauola}
\hrefCMSnoop {}{Z.~W{\c{a}}s, ``{TAUOLA the library for tau lepton decay, and
  KKMC/KORALB/KORALZ/... status report}'',} \textit{ {Nucl. Phys. B, Proc.
  Suppl.}} \textbf{ 98} (2001) 96,
  \href{http://dx.doi.org/10.1016/S0920-5632(01)01200-2}{\doi{10.1016/S0920-5632(01)01200-2}}.

\bibitem{geant4}
\hrefCMSnoop {}{{GEANT4} Collaboration, ``{GEANT4}---a simulation toolkit'',}
  \textit{ Nucl. Instrum. Meth. A} \textbf{ 506} (2003) 250,
  \href{http://dx.doi.org/10.1016/S0168-9002(03)01368-8}{\doi{10.1016/S0168-9002(03)01368-8}}.

\bibitem{Lai:2010vv}
H.-L. Lai\hrefCMSnoop {}{ {et~al.}, ``New parton distributions for collider
  physics'',} \textit{ Phys. Rev. D} \textbf{ 82} (2010) 074024,
  \href{http://dx.doi.org/10.1103/PhysRevD.82.074024}{\doi{10.1103/PhysRevD.82.074024}},
\href{http://www.arXiv.org/abs/1007.2241}{\texttt{arXiv:1007.2241}}.

\bibitem{CTEQpaper}
J.~Pumplin\hrefCMSnoop {}{ {et~al.}, ``{New generation of parton distributions
  with uncertainties from global QCD analysis}'',} \textit{ JHEP} \textbf{ 07}
  (2002) 012,
  \href{http://dx.doi.org/10.1088/1126-6708/2002/07/012}{\doi{10.1088/1126-6708/2002/07/012}},
  \href{http://www.arXiv.org/abs/hep-ph/0201195}{\texttt{arXiv:hep-ph/0201195}}.

\bibitem{Li:2012wna}
\hrefCMSnoop {}{Y.~Li and F.~Petriello, ``{Combining QCD and electroweak
  corrections to dilepton production in FEWZ}'',} \textit{ Phys. Rev. D}
  \textbf{ 86} (2012) 094034,
  \href{http://dx.doi.org/10.1103/PhysRevD.86.094034}{\doi{10.1103/PhysRevD.86.094034}},
\href{http://www.arXiv.org/abs/1208.5967}{\texttt{arXiv:1208.5967}}.

\bibitem{Czakon:2013goa}
\hrefCMSnoop {}{M.~Czakon, P.~Fiedler, and A.~Mitov, ``{Total Top-Quark
  Pair-Production Cross Section at Hadron Colliders Through
  $O(\alpha^4_S)$}'',} \textit{ Phys. Rev. Lett.} \textbf{ 110} (2013) 252004,
  \href{http://dx.doi.org/10.1103/PhysRevLett.110.252004}{\doi{10.1103/PhysRevLett.110.252004}},
\href{http://www.arXiv.org/abs/1303.6254}{\texttt{arXiv:1303.6254}}.

\bibitem{Brein:2012ne}
\hrefCMSnoop {}{O.~Brein, R.~V. Harlander, and T.~J.~E. Zirke,
  ``{vh@nnlo---Higgs Strahlung at hadron colliders}'',} \textit{ Comput. Phys.
  Commun.} \textbf{ 184} (2013) 998,
  \href{http://dx.doi.org/10.1016/j.cpc.2012.11.002}{\doi{10.1016/j.cpc.2012.11.002}},
\href{http://www.arXiv.org/abs/1210.5347}{\texttt{arXiv:1210.5347}}.

\bibitem{Kidonakis:2008mu}
\hrefCMSnoop {}{N.~Kidonakis and R.~Vogt, ``Theoretical top quark cross section
  at the {Tevatron and the LHC}'',} \textit{ Phys. Rev. D} \textbf{ 78} (2008)
  074005,
  \href{http://dx.doi.org/10.1103/PhysRevD.78.074005}{\doi{10.1103/PhysRevD.78.074005}},
\href{http://www.arXiv.org/abs/0805.3844}{\texttt{arXiv:0805.3844}}.

\bibitem{Campbell:2011bn}
\hrefCMSnoop {}{J.~M. Campbell, R.~K. Ellis, and C.~Williams, ``{Vector boson
  pair production at the LHC}'',} \textit{ JHEP} \textbf{ 07} (2011) 018,
  \href{http://dx.doi.org/10.1007/JHEP07(2011)018}{\doi{10.1007/JHEP07(2011)018}},
\href{http://www.arXiv.org/abs/1105.0020}{\texttt{arXiv:1105.0020}}.

\bibitem{ttWZ}
\hrefCMSnoop {}{M.~V. Garzelli, A.~Kardos, C.~G. Papadopoulos, and
  Z.~Tr{\'o}cs{\'a}nyi, ``{t${\rm \bar{t}}$${\rm W}^{\pm}$ and t${\rm
  \bar{t}}$Z hadroproduction at NLO accuracy in QCD with Parton Shower and
  Hadronization effects}'',} \textit{ JHEP} \textbf{ 11} (2012) 056,
  \href{http://dx.doi.org/10.1007/JHEP11(2012)056}{\doi{10.1007/JHEP11(2012)056}},
\href{http://www.arXiv.org/abs/1208.2665}{\texttt{arXiv:1208.2665}}.

\bibitem{ttW}
\hrefCMSnoop {}{J.~M. Campbell and R.~K. Ellis, ``{${\rm t \bar{t} W^{\pm}}$
  production and decay at NLO}'',} \textit{ JHEP} \textbf{ 07} (2012) 052,
  \href{http://dx.doi.org/10.1007/JHEP07(2012)052}{\doi{10.1007/JHEP07(2012)052}},
\href{http://www.arXiv.org/abs/1204.5678}{\texttt{arXiv:1204.5678}}.

\bibitem{Frederix:2011zi}
R.~Frederix\hrefCMSnoop {}{ {et~al.}, ``{Scalar and pseudoscalar Higgs
  production in association with a top-antitop pair}'',} \textit{ Phys. Lett.
  B} \textbf{ 701} (2011) 427,
  \href{http://dx.doi.org/10.1016/j.physletb.2011.06.012}{\doi{10.1016/j.physletb.2011.06.012}},
\href{http://www.arXiv.org/abs/1104.5613}{\texttt{arXiv:1104.5613}}.

\bibitem{Garzelli:2011vp}
\hrefCMSnoop {}{M.~V. Garzelli, A.~Kardos, C.~G. Papadopoulos, and
  Z.~Tr{\'o}cs{\'a}nyi, ``{Standard Model Higgs boson production in association
  with a top anti-top pair at NLO with parton showering}'',} \textit{ Europhys.
  Lett.} \textbf{ 96} (2011) 11001,
  \href{http://dx.doi.org/10.1209/0295-5075/96/11001}{\doi{10.1209/0295-5075/96/11001}},
\href{http://www.arXiv.org/abs/1108.0387}{\texttt{arXiv:1108.0387}}.

\bibitem{Alwall:2014hca}
J.~Alwall\hrefCMSnoop {}{ {et~al.}, ``{The automated computation of tree-level
  and next-to-leading order differential cross sections, and their matching to
  parton shower simulations}'',} \textit{ JHEP} \textbf{ 07} (2014) 079,
  \href{http://dx.doi.org/10.1007/JHEP07(2014)079}{\doi{10.1007/JHEP07(2014)079}},
\href{http://www.arXiv.org/abs/1405.0301}{\texttt{arXiv:1405.0301}}.

\bibitem{CMS-PAS-TOP-12-028}
\hrefCMSnoop {}{{CMS Collaboration}, ``{Measurement of differential top-quark
  pair production cross sections in pp colisions at $\sqrt{s}=7$ TeV}'',}
  \textit{ Eur. Phys. J. C} \textbf{ 73} (2013) 2339,
  \href{http://dx.doi.org/10.1140/epjc/s10052-013-2339-4}{\doi{10.1140/epjc/s10052-013-2339-4}},
\href{http://www.arXiv.org/abs/1211.2220}{\texttt{arXiv:1211.2220}}.

\bibitem{Khachatryan:2014uva}
\hrefCMSnoop {}{{CMS Collaboration}, ``{Differential cross section measurements
  for the production of a W boson in association with jets in proton-proton
  collisions at $\sqrt s=7$ TeV}'',} \textit{ Phys. Lett. B} \textbf{ 741}
  (2015) 12,
  \href{http://dx.doi.org/10.1016/j.physletb.2014.12.003}{\doi{10.1016/j.physletb.2014.12.003}},
\href{http://www.arXiv.org/abs/1406.7533}{\texttt{arXiv:1406.7533}}.

\bibitem{Punzi:2003bu}
\hrefCMSnoop {}{G.~Punzi, ``{Sensitivity of searches for new signals and its
  optimization}'',} in \textit{ Statistical problems in particle physics,
  astrophysics, and cosmology (PHYSTAT2003)}, L.~Lyons, R.~P. Mount, and
  R.~Reitmeyer, eds.
\newblock 2003.
\newblock
  \href{http://www.arXiv.org/abs/physics/0308063}{\texttt{arXiv:physics/0308063}}.

\bibitem{Chatrchyan:2012bra}
\hrefCMSnoop {}{{CMS Collaboration}, ``{Measurement of the t$\bar{\rm t}$
  production cross section in the dilepton channel in pp collisions at
  $\sqrt{s}=7$ TeV}'',} \textit{ JHEP} \textbf{ 11} (2012) 067,
  \href{http://dx.doi.org/10.1007/JHEP11(2012)067}{\doi{10.1007/JHEP11(2012)067}},
  \href{http://www.arXiv.org/abs/1208.2671}{\texttt{arXiv:1208.2671}}.

\bibitem{Chatrchyan:2013faa}
\hrefCMSnoop {}{{CMS Collaboration}, ``{Measurement of the ${\rm t \bar{t}}$
  production cross section in the dilepton channel in pp collisions at
  $\sqrt{s}$ = 8 TeV}'',} \textit{ JHEP} \textbf{ 02} (2014) 024,
  \href{http://dx.doi.org/10.1007/JHEP02(2014)024}{\doi{10.1007/JHEP02(2014)024}},
\href{http://www.arXiv.org/abs/1312.7582}{\texttt{arXiv:1312.7582}}.

\bibitem{CMS-PAS-LUM-13-001}
\href {http://cdsweb.cern.ch/record/1598864}{{CMS Collaboration}, ``{CMS
  Luminosity Based on Pixel Cluster Counting - Summer 2013 Update}'',} CMS
  Physics Analysis Summary CMS-PAS-LUM-13-001, 2013.

\bibitem{TagAndProbe}
\hrefCMSnoop {}{{CMS Collaboration}, ``{Measurement of the inclusive W and Z
  production cross sections in pp collisions at $\sqrt{s}=7$ TeV}'',} \textit{
  JHEP} \textbf{ 10} (2011) 132,
  \href{http://dx.doi.org/10.1007/JHEP10(2011)132}{\doi{10.1007/JHEP10(2011)132}},
\href{http://www.arXiv.org/abs/1107.4789}{\texttt{arXiv:1107.4789}}.

\bibitem{Chatrchyan:2012nj}
\hrefCMSnoop {}{{CMS Collaboration}, ``{Measurement of the inelastic
  proton-proton cross section at $\sqrt{s}=7$ TeV}'',} \textit{ Phys. Lett. B}
  \textbf{ 722} (2013) 5,
  \href{http://dx.doi.org/10.1016/j.physletb.2013.03.024}{\doi{10.1016/j.physletb.2013.03.024}},
\href{http://www.arXiv.org/abs/1210.6718}{\texttt{arXiv:1210.6718}}.

\bibitem{Alekhin:2011sk}
\hrefCMSnoop {}{S.~Alekhin {et~al.}, ``{The PDF4LHC Working Group Interim
  Report}'',} (2011).
\href{http://www.arXiv.org/abs/1101.0536}{\texttt{arXiv:1101.0536}}.

\bibitem{PDF4LHC}
M.~Botje\hrefCMSnoop {}{ {et~al.}, ``{The PDF4LHC Working Group Interim
  Recommendations}'',} (2011).
\href{http://www.arXiv.org/abs/1101.0538}{\texttt{arXiv:1101.0538}}.

\bibitem{Bourilkov}
\hrefCMSnoop {}{D.~Bourilkov, R.~C. Group, and M.~R. Whalley, ``{LHAPDF: PDF
  use from the Tevatron to the LHC}'',} (2006).
  \href{http://www.arXiv.org/abs/hep-ph/0605240}{\texttt{arXiv:hep-ph/0605240}}.

\bibitem{theta}
\hrefCMSnoop {}{J.~Ott}.
\newblock \url{http://www.theta-framework.org/}.

\end{thebibliography}\endgroup
\cleardoublepage \appendix\section{The CMS Collaboration \label{app:collab}}\begin{sloppypar}\hyphenpenalty=5000\widowpenalty=500\clubpenalty=5000\textbf{Yerevan Physics Institute,  Yerevan,  Armenia}\\*[0pt]
V.~Khachatryan, A.M.~Sirunyan, A.~Tumasyan
\vskip\cmsinstskip
\textbf{Institut f\"{u}r Hochenergiephysik der OeAW,  Wien,  Austria}\\*[0pt]
W.~Adam, E.~Asilar, T.~Bergauer, J.~Brandstetter, E.~Brondolin, M.~Dragicevic, J.~Er\"{o}, M.~Flechl, M.~Friedl, R.~Fr\"{u}hwirth\cmsAuthorMark{1}, V.M.~Ghete, C.~Hartl, N.~H\"{o}rmann, J.~Hrubec, M.~Jeitler\cmsAuthorMark{1}, V.~Kn\"{u}nz, A.~K\"{o}nig, M.~Krammer\cmsAuthorMark{1}, I.~Kr\"{a}tschmer, D.~Liko, I.~Mikulec, D.~Rabady\cmsAuthorMark{2}, B.~Rahbaran, H.~Rohringer, J.~Schieck\cmsAuthorMark{1}, R.~Sch\"{o}fbeck, J.~Strauss, W.~Treberer-Treberspurg, W.~Waltenberger, C.-E.~Wulz\cmsAuthorMark{1}
% ...
\vskip\cmsinstskip
% ...
1:~~Also at Vienna University of Technology, Vienna, Austria\\
2:~~Also at CERN, European Organization for Nuclear Research, Geneva, Switzerland\\
% ...

\end{sloppypar}
\end{document}